\documentclass[aps,prd,superscriptaddress,nofootinbib,showpacs,letterpaper,eqsecnum,twocolumn]{revtex4-1}

\usepackage{subfigure}
\usepackage{verbatim}
\usepackage[usenames, dvipsnames]{color}
\usepackage[pdftex]{graphicx}
\usepackage[english]{babel}
\usepackage{amsmath,amssymb}
\usepackage[colorlinks=true, citecolor=blue, linkcolor=magenta]{hyperref}
\usepackage{aas_macros}

\usepackage[T1]{fontenc}

\newcommand{\be}{\begin{equation}}
\newcommand{\ee}{\end{equation}}

\newcommand{\Rmnum}[1]{\expandafter\@slowromancap\romannumeral #1@}

\newcommand{\Eref}[1]{Eq.~\ref{#1}}
\newcommand{\eref}[1]{Eq.~\ref{#1}}

\begin{document}

\title{Towards the Laboratory Search for Space-Time Dissipation}
\author{Huan Yang}
\affiliation{Perimeter Institute for Theoretical Physics, Waterloo, Ontario N2L2Y5, Canada}
\affiliation{Institute for Quantum Computing, University of Waterloo, Waterloo, Ontario N2L3G1, Canada}
\email{hyang@perimeterinstitute.ca}
\author{Larry R.\ Price}
\author{Nicol\'as D.\ Smith}
\affiliation{Division of Physics, Math, and Astronomy, California Institute of Technology, Pasadena, CA 91125, USA}
\author{Haixing Miao}
\affiliation{School of Physics and Astronomy, University of Birmingham, Birmingham, B15 2TT, UK}
\author{Yanbei Chen}
\affiliation{Theoretical Astrophysics 350-17, California Institute of Technology, Pasadena, CA 91125, USA}
\author{Rana X.\ Adhikari}
\affiliation{Division of Physics, Math, and Astronomy, California Institute of Technology, Pasadena, CA 91125, USA}

\begin{abstract}
It has been speculated that gravity could be an emergent phenomenon, with classical general relativity as an effective, macroscopic theory, valid only for classical systems at large temporal and spatial scales.
As in classical continuum dynamics, the existence of underlying microscopic degrees of freedom may lead to macroscopic dissipative behaviors.  With the hope that such dissipative behaviors of gravity could be revealed by carefully designed experiments in the laboratory, we consider a phenomenological model that adds dissipations to the gravitational field, much similar to frictions  in solids and fluids.
Constraints to such dissipative behavior can already be imposed by astrophysical observations and existing experiments, but mostly in lower frequencies.  We propose a series of experiments working in higher frequency regimes, which may potentially put more stringent bounds on these models.  
\end{abstract}

\pacs{04.80.Nn, 95.55.Ym, 07.60.Ly}

\maketitle

\section{Introduction}
\label{sec:Intro}

Gravity has not yet been unified with other fundamental forces, which have been put into the Standard Model of particle physics.   General relativity,
currently the most successful theory for gravity, describes gravity as the classical geometry of space-time, which interacts only with the classical energy-momentum content of matter.  No fully consistent quantum theory of gravity has been found --- nor is it completely clear how quantum gravity can be experimentally probed.

%At this moment,
%gravity is the only fundamental force not described by the Standard Model of particle
%physics; yet there is currently no fully consistent quantum theory of gravity.

General relativity has been systematically verified with increasing accuracy
in the classical domain, with more opportunities arising from higher precision laboratory experiments and gravitational-wave detection~\cite{will2006confrontation,berti2015testing}.  On the other hand, having achieved great success in the microscopic world, quantum mechanics are now going to be tested in systems involving macroscopic mechanical objects, where gravity may start playing a role~\cite{marshall2003towards, romero2011large,pikovski2012probing,kaltenbaek2015macroscopic,kaltenbaek2012macroscopic,nimmrichter2013macroscopicity,yang2013macroscopic,nimmrichter2014optomechanical,diosi2015testing,chen2013macroscopic,kafri2015bounds,pfister2015understanding}. 

 In this paper, we shall still start off as testing general relativity, namely, testing the speculation that  general relativity
is not a fundamental theory, but an emergent one that appears to take place
only in macroscopic spatial and temporal scales.
In this work, we propose an experimental strategy to search for evidence of emergent gravity. We will soon see that the tests here will be intimately connected to tests of quantum mechanics, in systems where gravity plays a role.

%\subsection{The Case for Emergent Gravity}
%\label{sec:emergent_case}

Emergent gravity has been motivated from at least two types of
reasoning~\cite{hu2012emergence}. The {\it hydrodynamical} point of view, proposed by
Sakharov~\cite{Sakharov:1967pk,hu1996general}, argued that Einsteinian gravity can
naturally arise at large scales due to the dependence of vacuum energy (of non-gravitational fields)
on spacetime geometry.  The {\it thermodynamical} point of view focuses
on the role played by horizons, initially from the fact that the area of an event horizon
never decreases~\cite{Bardeen:1973gs} and the creation of particles by black
holes~\cite{Hawking:1974sw}.  Later, the work of Jacobson made a more general extension to
Rindler horizons of particles that can exist anywhere in a spacetime~\cite{Jacobson:1995ab}, and
made the connection between Einstein's equation and an equilibrium equation of state.   
%
%One mechanism  as the low-energy and
%semiclassical limit of a yet-unknown quantum gravity theory, may also be viewed
%as emergent.  
%
Recent work further considered the microscopic origin
of the horizon entropy --- motivated by the holographic principle~\cite{Verlinde:2010hp} and
by loop quantum gravity~\cite{Smolin:2010kk}.

A macroscopic dynamics emerging from an averaging over microscopic dynamics often
has two features: (i) {\it fluctuations} can arise from the imperfect averaging
over the microscopic degrees of freedom, (ii) the macroscopic equations of motion
can be non-conservative (often time asymmetric, or {\it dissipative}) due to excitations
of the microscopic degrees of freedom that do not instantaneously couple back
to the macroscopic dynamics. An instructive  example is the emergence of classical continuum dynamics from molecular
dynamics and (quantum) atomic physics.  The macroscopic properties of solids and liquids,
for example: (i) density, elastic moduli, compressibility, as well as (ii)
thermal conductivity, friction coefficient, viscosity all in fact emerge from the
microphysics of molecules.  Because fundamental physical laws are usually
considered to be time symmetric, those properties in group (ii), which are
associated with time asymmetric physical processes,  are direct signatures of the
underlying molecular dynamics.
In fact, physical processes associated with these properties had been discovered and
well characterized long before the underlying microphysics was well understood --- which
took place after more direct observations of molecular structure became feasible.
The above analogy generates hope that even though the microphysics that
underlies gravity may well be taking place at very high energies and very
small spatial and temporal scales, astrophysical observations and
laboratory experiments in the currently accessible regime might
be able to reveal their existence.

%\subsection{Testing Gravity versus Testing Quantum Mechanics}

Even without specifically considering the concept of emergence, testing
the validity of general relativity has long been a major research effort~\cite{will2006confrontation,berti2015testing}.  However,
most of these tests were performed in astrophysical or cosmological settings, for
example in the solar system~\cite{Reynaud:2008ts, Tury:2010} and in relativistic
binaries~\cite{Taylor:1994RMP, Kramer:2006cl}.
In the context of laboratory tests, deviation from Newton's inverse square law
at small distances has been proposed as a possible signature of models with
extra dimensions~\cite{Beane:1997kl, Randall:1999eq, ArkaniHamed:1998gb}. Experiments
have been performed to put bounds on such deviations~\cite{Jens:Grav2005}.
Experimental bounds on Lorentz invariance (of many kinds) have also evolved
substantially~\cite{PhysRevLett.4.176, Drever:1961cl, Tasson:2014dv, lrr-2005-5}
since the days of the Michelson-Morley experiment and now tightly constrain
several alternative gravity theories~\cite{Yuta:PRD13, Herrmann:2005tp}.  All such tests focus on low frequencies --- since the theories being tested for mainly have consequences in these regimes.  In this work, we shall consider families of theories  motivated by emergence, in particular those that have signatures toward higher frequencies and smaller length scales. We will consider a variant of  {\it non-local} models~\cite{hehl2009nonlocal,mashhoon2014nonlocal}, which were indeed motivated by space-time having ``memories'' in the past.  In these models, the 
gravitational field lags the motion of the source, even in the near zone.
We will  focus on tests on spatial and temporary scales accessible in the lab. 
%On the other hand, such phenomenological  models might be thought as low-energy effective theories of  more fundamental theories that include the unknown  microscopic degrees of freedom.

From a different starting points,  experiments have also been proposed for testing quantum mechanics involving macroscopic objects~\cite{marshall2003towards, romero2011large,pikovski2012probing,kaltenbaek2015macroscopic,kaltenbaek2012macroscopic,nimmrichter2013macroscopicity,yang2013macroscopic,nimmrichter2014optomechanical,diosi2015testing,chen2013macroscopic,kafri2015bounds,pfister2015understanding}.   Some of the theoretical modifications of quantum mechanics were motivated by gravity~\cite{diosi1984gravitation,penrose1998quantum,pikovski2012probing,yang2013macroscopic}, some are motivated by the determinism of the classical world~\cite{ghirardi1986unified,ghirardi1990markov}, some both~\cite{diosi1984gravitation,penrose1998quantum}.  Most of these experiments focused on the stochastic aspect of the modifications, although dynamical effects were also speculated~\cite{pikovski2012probing,diosi2014gravity,Diosi:2014a}.

In this work, we will recognize that our variant of non-local gravity~\cite{hehl2009nonlocal,mashhoon2014nonlocal} and the Diosi-Penrose (DP) gravity-induced decoherence~\cite{diosi1984gravitation,penrose1998quantum} can be unified --- both can arise from an infinite family of fields that couple with the energy-momentum content of space-time: non-local gravity as the dynamical consequence of the coupling, while DP decoherence as the stochastic back action.  Such a connection has been hinted by Diosi~\cite{diosi2014gravity,Diosi:2014a}, but we shall make a more general argument.   We will then explore possible experimental configurations that will test this unified model. 

This paper is organized as follows:  In Section~\ref{sec:Dissipation}, we
briefly explain how dissipation might be
incorporated theoretically, and set up a phenomenological model to test in the weak-field
regime, which is parametrized by phase-lag $\phi(\omega)$ (see the discussion in Sec.\ref{sec:PhenMod}), similar to the loss-angle used to characterize material losses \cite{Ting:Brownian2012}.  In
Section~\ref{sec:Constraints}, we briefly discuss how astrophysical processes
might be used to constrain $\phi(\omega)$, and argue that there exists a new
regime that is particular suitable for lab experiments. In Sec.~\ref{sec:Experiments},
we propose experimental strategies and explore their potential performance.

\section{Dissipation in Space-Time}
\label{sec:Dissipation}

%In this section, we set up strawman theories to fit in our intuition about spacetime dissipation. In particular, we seek out models in which the near-zone gravitational field lags relative to an oscillating source. We expect that such phase lags give rise to  dissipative equations of motion for moving masses. As our motivation is %based on macroscopic signatures that can be tested astrophysically or in lab experiments, these strawman models are {\it phenomenological}, in the sense that the unknown``environmental" or microscopic degrees of freedom no longer enters the dynamics, but rather affects the parameters of these models. It is also %possible that different micro-physical theories 
%lead to similar phenomenological models at low energies and large spatial scales.

%We note that most alternative gravity theories are formulated in terms of action principles, and are therefore necessarily conservative.  Dissipation only appears when we impose out-going boundary conditions.  In Sec.~\ref{subsec:existing}, we shall consider two classes of alternative theories which possess an energy loss %that creates an apparent dissipation. We explain how this differs from the type of dissipation we consider here, and therefore cannot be easily tested using the type of experiments we shall consider.
 %In Sec.~\ref{sec:PhenMod}, we shall introduce a phenomenological model with dissipation, and then study its dynamics in the linear regime.

In this section, we set up a strawman theory to fit in our intuition about spacetime dissipation. In particular, we would like to explore dissipative modifications to general relativity where the dynamics of spacetime is changed. This could be due to tracing out unknown ``environmental" or microscopic degrees of freedom, but here we focus only on a effective theory for the spacetime itself. In this sense, these strawman models are {\it phenomenological}, as the unknown microphysics no longer enters the dynamics, but rather affects the parameters of these models. It is also possible that different micro-physical theories 
lead to similar phenomenological models at low energies and large spatial scales.

In addition, we expect dissipative effect leads to a distinctive experimental signature in the linearized gravity regime: Newton's constant will effectively become complex in the frequency domain, so that for an oscillatory source, the phase of the Newtonian gravitational potential will lag behind that of the oscillatory of the motion, with a finite phase difference, {\it even in the near zone}.
We shall discuss the phenomenology of such models, illustrating how they differ from traditional modified gravity models.

\subsection{Non-local Einstein's Equation and its relation to existing modifications to GR}\label{sec:PhenMod}

Let us assume that the presence of unknown microphysics introduces a delayed-response for the spacetime geometry with respect to its matter source. This delayed-response effectively makes the  Einstein equations to be non-local, which may be written as
\begin{equation}\label{eqtimeg}
G_{\mu\nu}(t,\mathbf{x})=8\pi \int^{+\infty}_{-\infty}K_{\mu\nu\alpha\beta}(t-t')T^{\alpha\beta}(t',\mathbf{x})dt'.
\end{equation}
This is a generalization of non-local gravity models proposed by Hehl and Mashhoon~\cite{hehl2009nonlocal,mashhoon2014nonlocal}.  The form of Eq.~\eqref{eqtimeg} is already written into a 3+1 form, which indicates that the Einstein tensor at one space-time event is determined as a weighted average over the worldlines of a family of fiducial observers.  The existence of these special observers implied a preferred frame --- as a consequence of the  unspecified miscrophysics, similar to the effect of extra degrees of freedom in other modified gravity theories --- for example the Einstein-aether theory \cite{Jacobson2007}, where the configuration of the aether field effectively defines a preferred frame.  \footnote{Some other approaches to nonlocal gravity include \cite{Nicolini:2005vd, Modesto:2010uh} non-commutative
geometry effects.} 

The filter  $K_{\mu\nu\alpha\beta}$ is a {\it bi-tensor}, and  parametrizes possible delayed-responses.  It is symmetric in its first and second pairs of indices (i.e. $K_{\mu\nu\alpha\beta}$ = $K_{\nu\mu\alpha\beta}$ = $K_{\mu\nu\beta\alpha}$), in order to respect the symmetry of $G_{\mu\nu}$ and $T_{\mu\nu}$.  Since we have already chosen a preferred set of observers for the non-local integral, let us also perform a 3+1 decomposition of different components of the tensor $K_{\mu\nu\alpha\beta}$.  To illustrate properties of such models, we choose a filter function  of
\begin{align}\label{eqspecialk}
K_{\mu\nu\alpha\beta}(t-t') &=g_{\mu\alpha} g_{\nu\beta}\delta(t-t') \nonumber\\
&+
 [K(t-t')-\delta(t-t')]g_{\alpha 0} g_{\beta 0} {g_{\mu }}^0 {g_{\nu }}^0
\end{align}
so that only the $00$ component of the Einstein equations is changed. In the vacuum case we recover the source-free Einstein equations $G_{\mu\nu}=0$, which suggests that the propagation of gravitational waves is unchanged in this modified gravity theory. Therefore in order to experimentally test its effect, we need to study the cases where the stress energy tensor of matter is nonzero. As explained in Sec.~\ref{sec:Bianchi}, the matter equation of motion is modified, which can be derived from Eq.~\ref{eqtimeg} and the Bianchi identity.

In addition, Eq.~\ref{eqtimeg} shows that the Einstein tensor $G_{\mu\nu}$ depends non-locally on the stress-energy tensor $T_{\mu\nu}$; in other words, space-time curvature depends non-locally on its energy-momentum content.   In order for this dependence  to be causal, we must impose
\begin{equation}
\label{condK}
K(t)=0\,,\quad t<0\,.
\end{equation}
Let us write the Fourier transformation of $K$ as
\begin{equation}
\int^\infty_{-\infty} dt  \,e^{i \omega t} \, K(t) \equiv e^{i \phi(\omega)}\,,
\end{equation}
where $\phi(\omega)$ is expected to be small. If $\phi(\omega)$ is a rational function,
condition \eref{condK} can be achieved by requiring $\phi(\omega)$ to have no poles on the upper half complex plane.  Furthermore, in order to keep $G_{\mu\nu}$ and $T_{\mu\nu}$ real-valued in the time domain, we need to impose $K(t) \in \mathbf{R}$, or
\begin{equation}
\phi(\omega) = -\phi^*(-\omega) \,,\quad \omega\in\mathbf{R}\,.
\end{equation}
In particular, $\phi(0) \in \mathbf{R}$, which can be absorbed  into the definition of the Newton's constant, leading to
\begin{equation}
\label{eqphi0}
\phi(0) =0\,.
\end{equation}

Furthermore, in order for $\phi$ to represent a phase lag (instead of a  variation in the magnitude of the Newton's constant), we shall require $\phi(\omega)$ to be real-valued for low frequencies --- although in general requiring $\phi(\omega)$ to be real-valued for all frequencies may conflict with the requirement that $K$ be causal.  One prescription is to choose
\begin{equation}
\label{phisimple}
e^{i \phi(\omega)} =\frac{1}{1-i \omega \tau_*}, \quad {\rm or} \quad \phi(\omega) \approx \omega \tau_* \,,\quad \tau_* >0
\end{equation}
which leads to
\begin{equation}\label{eqpkernel}
K(t) =\frac{1}{\tau_*}e^{-t/\tau_*} \Theta(t)\,,
\end{equation}
with $\Theta$ the Heaviside step function.  With this filter function, the space-time geometry has a (short) response time of $\tau_*$.

We assume that the dynamics of microscopic degrees of freedom only takes place at high frequencies, while in low frequencies accessible by astrophysical processes and our experiments, $\phi$ can be simply Taylor expanded as
\begin{equation}\label{eqpexp}
\phi(\omega) = \omega \phi'(0)  +\frac{\omega^2}{2} \phi''(0) +\ldots\,.
\end{equation}
Since the leading order term of the above expansion is the same as \Eref{phisimple}, we can identify
\begin{equation}
\tau_* \approx \phi'(0) \approx \frac{\phi(\omega)}{\omega}\,,
\end{equation}
which all have the physical meaning of time-lag.

%\textcolor{red}{HERE WE SHOULD MENTION THAT THIS MODEL HAS NOT BEEN STUDIED BEFORE AS MOST MODIFIED GRAVITY MODELS HAS A SMALL NUMBER OF ADDITIONAL FIELDS, THEREFORE DO NOT HAVE DISSIPATION OF THIS TYPE?}

%\textcolor{red}{What about models in which particles's wave packets decay over propagation?}

\subsection{Linearized Einstein's Equation and its solution for periodic source}

In the case where gravity is weak, we can simplify Eq.~\ref{eqtimeg} by considering a perturbed flat metric $g_{\mu\nu} = \eta_{\mu\nu}+h_{\mu\nu}$, with $|h| \ll 1$.
Within such limit, Eq.~\ref{eqtimeg} can be linearized, while the only different component from linearized Einstein's equation is (C.f. Eq.~\ref{eqtimeg} and Eq.~\ref{eqspecialk})
\begin{equation}\label{eqwl}
\left[-\omega^2 +\nabla^2\right]\bar h^{00} = -16\pi e^{i \phi(\omega)}T^{00}\,.
\end{equation}
Here $\bar h_{\mu\nu}$ is the trace-reversed spacetime metric perturbation,
\begin{eqnarray}
\label{gauge}
\bar{h}_{\mu\nu} &=&h_{\mu\nu}-\frac{h }{2}\eta_{\mu\nu}\,,
\end{eqnarray}
which satisfies the gauge condition (with $4$-dimesional covariant derivative)
\begin{equation}\label{eqgauge}
\nabla^\mu \bar h_{\mu\nu}=0\,.
\end{equation}
Note that in order to obtain Eq.~\eqref{eqwl} we have performed a 3+1 split of coordinates, transformed $t$ into the frequency domain, and used $\mathbf{x}$ to denote spatial coordinates.
As discussed in Sec.~\ref{sec:PhenMod}, $\phi(\omega)$
is a real-valued function characterizing phase-lag as functions of frequency, and $T_{\mu\nu}(\omega,\mathbf{x})$ is the stress-energy tensor of the source. The gauge condition described in Eq.~(\ref{eqgauge}) considerably simplifies the form of linearized Einstein's equation, and it is still compatible with dissipative-gravity modifications, as the left hand side of Eq.~\eqref{eqtimeg} is unmodified. 
%As demonstrated later, the Bianchi identity (${G^{\mu\nu}}_{;\nu}\equiv 0$) in this phenomenological model no longer implies the divergence-free equations of motion (${T^{\mu\nu}}_{;\nu}=0$). The new equations of motion are easily obtained by combining the Bianchi identity with Eq.~(\ref{eqtimeg}).

We now solve the wave-equation Eq.~\eqref{eqwl}
 and obtain a retarded solution
\begin{equation}\label{potentialbar}
{\bar{h}}_{00}(f,\mathbf{x}) = 4 \int d^3{\bf x}'\, \frac{{T}_{00}(f,\mathbf{x}')}{|\mathbf{x}-\mathbf{x}'|} e^{i \left[2\pi f  |\mathbf{x}-\mathbf{x}'|+\phi(\omega)\right]}\, .
\end{equation}
Suppose the  source is oscillating at a constant frequency $f_0=\omega_0/2\pi$. In the near zone ($2\pi f_0 R/c \ll 1$) we have
\begin{equation}
	\label{newt}
\bar{h}_{00}(t,\mathbf{x}) \approx 4 \int d^3\mathbf{x'}\, \frac{T_{00}(t,\mathbf{x}')}{|\mathbf{x}-\mathbf{x}'|} e^{i \phi(\omega_0)}.
\end{equation}
Again, the form of Eq.~(\ref{eqwl}) together with the requirement that $T_{\mu\nu}$ and $\bar h_{\mu\nu}$ be real-valued in the time domain, require that  $\phi(0) = 0$.  The power series expansion in Eq.~\ref{eqpexp}
then allows us to switch freely between $\phi$ and $\phi'$ at low frequencies. Physically $\phi$ carries the interpretations of a phase lag while $\phi'$ is a time lag.  We also note that this expansion does not necessarily hold at high frequencies, where microphysics of the underlying ``environment" may contribute to a very different $\phi$. For concreteness, subsequent sections detailing existing constraints and experimental sensitivity, which are supposedly targeting much lower frequencies comparing to underlying microscopic motions, will have final results phrased in terms of $\phi$.

To see how this model modifies near zone interactions, we work with the gravitational potential  $U_{\rm DG}$ of our model, defined as
\begin{eqnarray}
g_{00} &= \eta_{00} + h_{00}\\
&= -1 + 2U_{\rm DG}.
\end{eqnarray}
Using this with Eq.~\ref{newt} and Eq.~\ref{gauge}, for a periodic source moving at frequency $f_0$, we see that
\begin{equation}
U_{\rm DG} = \int d\mathbf{x}'\, \frac{\rho(t,\mathbf{x}')}{|\mathbf{x}-\mathbf{x}'|} e^{i \phi(\omega_0)},
\end{equation}
where $\rho(t,\mathbf{x}') = T_{00}(t,\mathbf{x}')$.  Note that this differs from the usual Newtonian potential, $U$, simply by a phase, i.e.,
$U_{\rm DG} = Ue^{i \phi(\omega_0)}$.

\subsection{Bianchi Identity and equations of motion in the Newtonian limit}\label{sec:Bianchi}

Now let us consider the equation of motion for matter based on Eq.~\eqref{eqtimeg}. Unlike last section where we focus on the metric in the wave-zone, here we are mainly interested in the matter motion in the Newtonian near-zone. The Bianchi identity ${G^{\mu\nu }}_{|\nu} = 0$ requires that
\begin{equation}\label{eqbianchi}
{G^{\mu\nu }}_{,\nu}+{\Gamma^{\mu}}_{\nu\alpha} G^{\alpha \nu}+{\Gamma^{\nu}}_{\nu\alpha} G^{\mu\alpha}=0\,,
\end{equation}
where the Einstein tensor $G^{\mu\nu}$ is to be replaced by the right hand side of Eq.~\ref{eqtimeg}, and the above equation becomes the modified equation of motion for matter. Because of the modification, the matter motion is generically non-conservative. As a simple example,  applying Eq.~\ref{eqbianchi} and Eq.~\ref{eqtimeg}  to a point mass in the Newtonian limit,  the equation of motion is just
\begin{align}
0\approx &\left [T^{i 0}(t,{\bf x}) \right ]_{,0}+{\Gamma^{i}}_{00} \int^\infty_{-\infty} dt' K(t-t') T^{00}(t', {\bf x}) \nonumber \\
\approx & m \left [ v^i(t) \right ]_{,0}-m \frac{1}{2} \partial^{i} h_{00} \int^{\infty}_{-\infty} dt' K(t-t')\,.
\end{align}
As we anticipate that the theory recovers the Einstein gravity in the static limit, $\int^\infty_{-\infty} K(t-t') dt'$ should be set to one. As a result, the point mass equation of motion becomes
\begin{equation}\label{eqpeom}
a^i(t) = \frac{f^i(t)}{m}\,,
\end{equation}
where $f^i$ is the Newtonian gravitational force acting on the point mass. As we can read from Eq.~\ref{eqtimeg} and discussion in later sections, the metric is affected not only by local matter stress-energy, but also contributions of the past light-cone. In particular,
\begin{equation}
\frac{f_i (t)}{m} = \int^\infty_{-\infty} d t' K(t-t') \Psi_{,i}(t',{\bf x(t')})\,,
\end{equation}
and here $\Psi$ is the Newtonian potential. Comparing the above equation with the point mass equation of motion, we find that
\begin{equation}\label{eqacce}
a_i (t) = \int^\infty_{-\infty} d t' K(t-t') \Psi_{,i}(t',{\bf x(t')})\,.
\end{equation}
This means that within the Newtonian limit, this dissipative gravity model leads to an  different equation of motion  from General Relativity.

\subsection{Relation between non-local gravity and spontaneous wave-function collapse}

%From a different starting points,  experiments have also been proposed for testing quantum mechanics involving macroscopic objects~\cite{marshall2003towards, romero2011large,pikovski2012probing,kaltenbaek2015macroscopic,kaltenbaek2012macroscopic,nimmrichter2013macroscopicity,yang2013macroscopic,nimmrichter2014optomechanical,diosi2015testing,chen2013macroscopic,kafri2015bounds,pfister2015understanding}.   Some of the theoretical modifications of quantum mechanics were motivated by gravity~\cite{diosi1984gravitation,penrose1998quantum,pikovski2012probing,yang2013macroscopic}, some are motivated by the determinism of the classical world~\cite{ghirardi1986unified,ghirardi1990markov}, some both~\cite{diosi1984gravitation,penrose1998quantum}.  Most of these experiments focused on the stochastic aspect of the modifications, although dynamical effects were also speculated~\cite{pikovski2012probing,diosi2014gravity,Diosi:2014a}.   
%

Let us turn to collapse models of quantum mechanics.  The initial motivation for these models were from the randomness of quantum-state reduction, and its incompatibility with classical determinism.  The assumption of these models are that quantum states of macroscopic objects spontaneously become localized due to an {\it intrinsic} collapse process~\cite{ghirardi1986unified,ghirardi1990markov,diosi1984gravitation,penrose1998quantum}.  As was later realized, these collapses can all be modeled in general as a continuous measurement on matter density~\cite{nimmrichter2013macroscopicity,nimmrichter2014optomechanical} --- with spatially distributed measuring devices that have different mutual correlations. 

In a continuous quantum-measurement process, back action takes the form of back-action noise, but also sometimes in the form of dissipation.  This has lead Diosi to further propose that the collapse models may also cause the Newtonian gravitational potential to have a delay $\tau_*$ when responding to matter density changes~\cite{Diosi:2014a}.  This is the same as Eq.~\eqref{eqpexp} as we consider the lowest-frequency contribution from non-local gravity.    Diosi argued that current experimental data can constrain $\tau_*$ to around 1\,ms. 
However, in this paper we shall remain flexible about the function form of $\phi(\omega)$.

Let us further argue that a more general $\phi(\omega)$ can also arise from the quantum-measurement consideration. More specifically, a continuous measurement process can be modeled as coupling the observable we need to measure with a field degree of freedom, which has an incoming state, and the out-going state contains information about that degree of freedom.   Quantum or thermal fluctuations in the incoming state provides the stochastic back-action, while information contained in the out-going state corresponds to dissipation. 

Now suppose we apply this to space-time geometry. If $h_{00}(t,\mathbf{x})$ is being continuously monitored, we need to introduce an extra dimension $\eta$, and an additional field $n$ --- for each $\mathbf{x}$ --- that propagates on the $t$-$\eta$ plane. The field couples to $h_{00}$ as $\eta$ approaches 0 --- it therefore gains information about $h_{00}$, and acts back to the dynamics of $h_{00}$.  In this way, depending on the propagation law of $n$ along the extra dimension, and the detailed way it is couple to $h_{00}$, arbitrary shapes of the phase delay $\phi(\omega)$ between $h_{00}$ and energy density can be constructed. 

In general, as we consider different $\mathbf{x}$, coupling to different components of the metric, as well as the above dependence on $\eta$, we can recover general non-local theories of gravity.

%We assume that the dynamics of microscopic degrees of freedom only takes place at high frequencies, while in low frequencies accessible by astrophysical processes and our experiments, $\phi$ can be simply Taylor expanded as
%\begin{equation}\label{eqpexp}
%\phi(\omega) = \omega \phi'(0)  +\frac{\omega^2}{2} \phi''(0) +\ldots\,.
%\end{equation}
%Since the leading order term of the above expansion is the same as \Eref{phisimple}, we can identify
%\begin{equation}
%\tau_* \approx \phi'(0) \approx \frac{\phi(\omega)}{\omega}\,,
%\end{equation}
%which all have the physical meaning of time-lag.

\section{Effects on one and two-body motions}

In this section, we shall discuss the dynamical effect of emergent gravity on one and two bodies.

\subsection{Effects on Single Body Motion}
\label{sec:singlebodyeffect}

Consider a spherical, homogeneous object (with mass $M$, radius $R$) moving with non-relativistic velocity $v$ relative to the preferred frame where Eq.~\eqref{eqtimeg} holds.  According to  Eq.~\eqref{eqacce}, there is a force generated by the ``past'' gravitational field of the same object. This self-gravitational force is dissipative for the object's motion, as it generates acceleration anti-parallel to the direction of motion:
\begin{equation}
{\bf a} = -\frac{G M}{R^3} {\bf v} \int^\infty_{0} \, t\, K(t)d t\,,
\end{equation}
where $G$ has been restored in this equation.

If we assume that the kernel function is described by Eq.~\eqref{eqpkernel}, the acceleration evaluates to
\begin{equation}
\label{eq:spacetimeviscocity}
{\bf a} = -\frac{G M}{R^3} \tau_* {\bf v} = -\frac{G M}{R^3} \phi'(0) {\bf v}\,.
\end{equation}
The phase-lagging feature of the nonlocal Einstein equation inevitably leads to a dissipative self-interacting gravitational field.

\subsection{Effects on Two-Body Motion}
\label{sec:GW}

In order to calculate effects on two-body interactions, we take a step back and look at the ``full'' theory.  Consider
\begin{align}
G_{00}(f,\bf{x}) &= 8\pi e^{i\phi(2\pi f)}T_{00}(f,\bf{x})\nonumber\\
						&\approx 8\pi[1+i\phi(2\pi f)]T_{00}(f,\bf{x}).
\end{align}
Writing the Fourier transform of $i\phi(\omega)$ as $\Phi(t)$ (which is $K(t)-\delta(t)$), the time-domain version becomes
\begin{equation}
G_{00}=8\pi\left[T_{00}+\int^0_{-\infty}\Phi(t-t')T_{00}(t')dt'\right].
\end{equation}
Assuming $\Phi(t-t')$ is a causal kernel implies $\Phi(t-t')=0$ for $t<t'$, allowing us to extend the limits of the integral to $\pm\infty$, resulting in our final expression
\begin{equation}\label{eqintkernal}
G_{00}=8\pi\left[T_{00}+\int^\infty_{-\infty}\Phi(t-t')T_{00}(t')dt'\right].
\end{equation}

 \begin{figure}[h]
	\includegraphics[width=0.8\columnwidth]{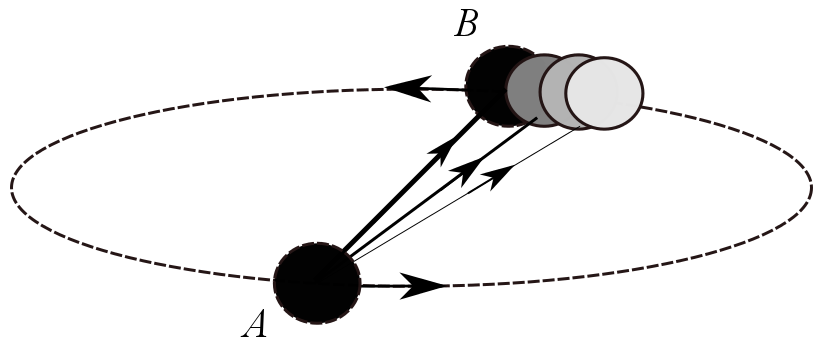}
\caption{Binary motion taking into account the dissipative gravity effect. In the Newtonian limit, star $A$ feels the instantaneous gravitational attraction from star $B$, as well as the contribution form ``past" gravitational field due to the dissipative gravity effect, according to Eq.~(\ref{eqintkernal}).}
\label{fig:Binaries}
\end{figure}

To estimate the effect in binary motion, we consider two equal mass stars A and B in orbit around each other as shown in Fig.~(\ref{fig:Binaries}).  In the Newtonian limit, star $A$ and $B$ feel the instantaneous attraction force from each other, which are orthogonal to their velocities. If we further take into account the dissipative gravity effect, as shown by Eq.~(\ref{eqintkernal}), the external gravitational field acting on star A (and vice versa for star B) contains a piece that originates from the past of star B. These ``past" gravitational fields exert a tangential force on star A's orbital motion, which in turn introduces additional orbital energy loss/gain (depending on the sign of $\phi(\omega)$) with respect to the gravitational radiation.  For example, if $\phi(\omega)>0$, although the gravitational field lags behind the source motion, the tangential force is along the star's direction of motion and  and  the binary motion actually gains energy according the positive phase lag. 
In this case we see that the rate of energy gain is given by
\begin{align}
\dot{E}_{\rm MG} &\sim   q \frac{M^2}{r^2}  \int^0_{-\infty}\Phi(t-t')\sin[2\pi f(t-t')]dt' \nonumber \\
&  \sim q (M f)^{4/3} \phi(2\pi f),
\label{eq:ecg}
\end{align}
where $q=m_1m_2/(m_1+m_2)^2$ is the symmetric mass ratio, the post-Newtonian parameter $v^2=(2\pi M f)^{2/3}$, and the
 subscript $MG$ stands for mutual gravity. As we  require that the standard Newtonian result to be recovered in the static limit, this constrains the DC phase-lag such that $\phi(f=0)=0$.  Therefore if the  frequency is sufficiently low,  on can Taylor-expand the phase and obtain
\begin{equation}
\dot{E}_{\rm MG}  \sim q M^{-1} (M f)^{7/3} \phi'_0.
\end{equation}

However, this is not the leading order dissipative gravity effect, which comes from the self-dissipative dragging force as shown in the previous section. In the binary system, such a damping mechanism dissipates energy at rate
\begin{equation}
\dot{E}_{\rm SG} = - \left ( \frac{m^2_1}{R^3_1}v_1+\frac{m^2_2}{R^3_2} v_2\right ) \phi'_0= - q M v \phi'_0 \left ( \frac{m_1}{R^3_1}+\frac{m_2}{R^3_2}\right ) ,
\end{equation}
which is about a factor of $r^3/R^3$ larger than $\dot{E}_{\rm MG}$.
To help further understand this result, recall that quadrupole radiation of gravitational waves has power
\begin{eqnarray}
\dot{E}_2 \sim -\frac{32}{\pi}q^2 (M f)^{10/3},
\end{eqnarray}
which is  one and a half post-Newtonian order higher than $E_{\rm MG}$, and consequently even smaller comparing to $E_{\rm SG}$!  A direct consequence of this observation is that binary pulsars should
provide the best observational constraints on the magnitude of $\phi$ at low frequencies, because of the high compactness of neutron stars.

\section{Existing Observational and Experimental Constraints}
\label{sec:Constraints}

\begin{table}
\centering
    \begin{tabular}{cccc}
  {Experiment}      & \begin{tabular}{c} Upper limit on \\ $\phi(2 \pi f_0)$ \end{tabular}& \textbf{$f_0$} (Hz)            & {Reference} \\ \hline \hline
    PSR J0737-3039               &      $7.9 \times 10^{-36}$                         &      $1.1\times10^{-4}$       &   \cite{2006Sci...314...97K}                \\ \hline
    PSR B1534+12               &       $ 4.3 \times 10^{-38}$                      &      $2.8\times10^{-5}$       &   \cite{2002ApJ...581..501S}                \\ \hline
    PSR B1913+16               &           $ 2.4 \times 10^{-39}$                   &      $3.6\times10^{-5}$       &   \cite{2010ApJ...722.1030W}                \\ \hline
    PSR B2127+11C               &                $1.6 \times 10^{-37}$                &      $3.5\times10^{-5}$       &   \cite{2006ApJ...644L.113J}                \\ \hline
    Earth Orbital Motion       &         $3\times 10^{-17}$                                               &        $3.2\times10^{-8}$     &         \cite{Stepehson1997}          \\ \hline
    Shapiro delay               &         $0.014$                                                &      $3.2\times10^{-8} $                                 &   \cite{Kopeikin:2001, Fomalont:2003, Shapiro:1966cp}                \\ \hline
    Torsion Pendulum       &  $0.1^\dagger$                                                          & $0.084$                   & \cite{Jens:Grav2007} \\ \hline
    Quadrupole Antenna       &  $0.1^\dagger$                                                      & $60.5$                   & \cite{Hirakawa:1980fh} \\ \hline
    Cantilever                     &  $1\times10^{-3}$                                   & 324.1                                  & \cite{Geraci:2008hp}         \\ \hline
    Rotor/Cantilever        &  0.1                                                         & 353                                     & \cite{Weld:Gas2008}         \\ \hline

    \end{tabular}
\caption{Constraints on $\phi$ derived from existing measurements at different frequencies.   Quantities with a ${}^\dagger$ superscript are estimated by
  assuming a $0.1$~rad uncertainty in the phase at the specified frequency. The phase lag in the Shapiro delay measurement is obtained by converting the constraint on the time-lag estimate on the same system.}
\label{tab:constraints}
\end{table}

\subsection{Binary pulsars}
According to our previous analysis, if we turn on the dissipative gravity effect, the change of fractional energy emission rate for a binary pulsar system is
\begin{align}
\frac{\Delta\dot{ E}}{\dot{E}} =  \frac{\dot{E}_{\rm SG}}{\dot{E}_2}&=\frac{(2\pi)^{2/3} \pi \phi'_0 M}{32q(Mf)^{3}} \left ( \frac{m_1}{R^3_1}+\frac{m_2}{R^3_2}\right ) \nonumber\\
&\approx \frac{(2\pi)^{2/3} \pi \phi'_0 M^2 }{32q(Mf)^{3} R^3} .
\end{align}
and, denoting $P_b$ as the period of the binary motion, we have
\begin{equation}
\frac{\Delta \dot{P_b}}{\dot{P_b}} =\frac{3}{2} \frac{\Delta \dot{E}}{\dot{E}}.
\end{equation}
Therefore the accuracy of period measurement constrains the magnitude of possible dissipative gravity effect. We can then use the fact that $\phi'_0 = \phi(2 \pi f_0)/(2 \pi f_0)$ to phrase our results in terms of the phase angle. The compactness of neutron stars $M_{1,2}/R_{1,2}$ can be determined from different models of nuclear equation of state, and here for simplicity the neutron star radius $R$ is estimated as $13 km$. Constraints from a representative sample of binary pulsars appear in Table~\ref{tab:constraints}.  We can see that pulser systems give really stringent upper bound on $\phi$ at low frequencies (recall the $1 ms$ constraint on $\tau_*$ in \cite{Diosi:2014a}), as nuclear densities are much higher than normal matter. On the other hand, it remains interesting to constrain $\phi$ at higher frequencies, using table-top experiments.

\subsection{Solar system test}

Solar system test of gravity theories generally involves measuring perihelion precession, spin precession measurement (such as ``Gravity Probe B") and test of the ``Weak Equivalence Principal". None of the above experiments seems to constrain a friction-type force well \cite{Hamilton2008}. We can nevertheless expect the period change after one orbital period of earth:
\begin{equation}
\frac{\Delta T}{T} = \frac{3}{2} \frac{\Delta E}{E} = 3 \frac{G M_E T}{R^3_E v} \phi'_0\,,
\end{equation}
where $M_E, R_E$ are the earth mass and average radius respectively. Plugging in the numbers, we find
\begin{equation}
\phi'_0 \sim2 \times 10^2 \frac{\Delta T}{T} s \sim 1.4 \times 10^{-10} s\,,
\end{equation}
where we have used the orbital period change of earth due to tidal friction $\sim 2.3 {\rm ms}/{\rm century}$ as an estimate \cite{Stepehson1997}. Obviously this is a much looser constraint than the binary pulsar tests, because  normal matter density is much lower than the nuclear matter density. As the earth orbital period is longer than the periods of binary pulsars, this test belongs to the low-frequency regime which is ruled out by binary pulsar tests.

\subsection{Shapiro Delay}
Shapiro delay is the time delay of light due to the local gravitational field of a nearby massive body~\cite{Demorest:2010bf, Shapiro:1966cp}. For a static source, our model predicts the same static gravitational field as general relativity, so there is no observable Shapiro delay effect.  However, as we have shown above, this is not the case for a moving source.

In the solar system, the time delay of light for a moving source has been measured in 2002~\cite{Kopeikin:2001, Fomalont:2003}, as Jupiter moved by the line of sight to quasar J0842+1835.  The timing sequence of the delay signal was applied to constrain the speed of gravity, which was measured to be consistent with the speed of light, with a relative uncertainty up to $20\%$. We notice that the phase-lag due to dissipative gravity, can be effectively translated to a shift in the ``speed of gravity" in this case. More specifically, if the closest distance between the trajectory of the light is $d$ and the velocity component of Jupiter moving towards the light trajectory is $v_d$, then the time lag is roughly bounded by
\begin{equation}
\frac{\phi' v_d}{d} \le 20\%\,, {\rm or} \quad \phi' \le \frac{d}{ 5 v_d} \,.
\end{equation}

According to the parameters presented in \cite{Kopeikin:2001, Fomalont:2003}, $v_d \sim 14 \,km/s$ and $d \sim \,10^6 km$, and consequently we obtain $\phi' \le 7\times 10^4 s$. Because the moving-body correction to the Shapiro delay is already a second order effect, in terms of $v_d/c$, this measurement a gives much weaker constraint on $\phi'$ than the test based on planetary motion.

\subsection{Near field experiments}
Although astrophysical tests provide the tightest constraints on our model, laboratory-scale experiments designed to test the nature of near-field gravity provide constraints at much higher frequencies.  Examples include torsion pendulua~\cite{Jens:Grav2007} and the quadruple antenna~\cite{Hirakawa:1980fh}, which look for deviations in the amplitude and do not report phase information.  In these cases, we assume the experiment has observed no phase lag greater than $.1$ radians.  Other experiments involving cantilevers~\cite{Geraci:2008hp} and improvements thereof~\cite{Weld:Gas2008} explicitly measure the phase angle.  Table~\ref{tab:constraints} summarizes these constraints.

The self-damping effect for moving bodies described in Sec.~\ref{sec:singlebodyeffect} has the consequence that there is an intrinsic damping force experienced by any harmonic oscillator involving moving masses. From Eq.~(\ref{eq:spacetimeviscocity}), and assuming a harmonic oscillator with natural angular frequency $\omega_0$, the mechanical quality factor is limited by
\begin{equation}
Q^{-1} \ge \frac{G \rho}{\omega_0}\tau_*(\omega_0) = \frac{G \rho}{\omega_0^2}\phi(\omega_0),
\end{equation}
where $\rho=M/R^3$, and we have allowed the gravitational delay, $\phi(\omega)$, to be frequency dependent. Here we see that the damping effect is greater for low frequency oscillators with dense masses. Depending on the value of $\phi$, this damping could exceed other mechanical dissipation effects. If this is not the case, the mechanical quality factor will only be slighly modified by gravitational self-damping and estimating the non-gravitational mechanical damping could lead to large systematic uncertainty in the estimation of the gravitational phase lag. For the oscillators described in Sec.~\ref{sec:Experiments}, we find that the contraint placed by direct measurement of the phase lag to be more stringent than the constraint set by measuring the quality factor.

\subsection{Gravitational wave experiments}
It is clear from Sec.~\ref{sec:GW} that our model has implications for gravitational wave detection, effecting the coalescence of compact binaries.  In general, the phase evolution of the waveforms will be modified and a careful study of the full implications of Eq.~(\ref{eqtimeg}) is required.  These issues will be explored in future studies.  However, for the model under consideration here, binary pulsars should provide a stronger constraint for low frequency motions.

\begin{figure}
\centering
\includegraphics[width=\columnwidth]{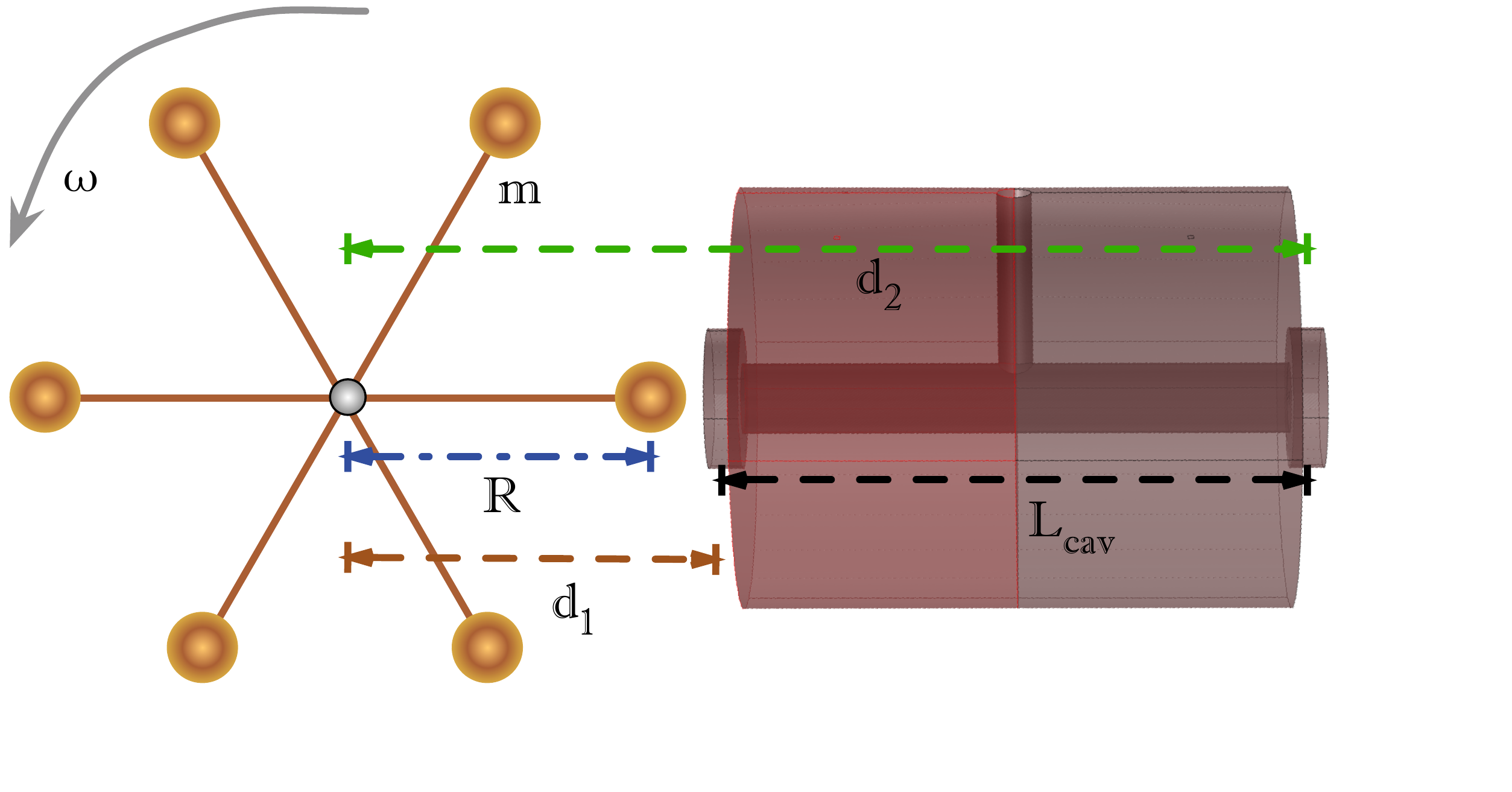}
  \caption{Plan view of a schematic diagram of the experiment. In this example,
  a rotor with 6 spokes
  and masses spins at an angular frequency $\omega$ near a rigid Fabry-Perot cavity
  with an angular eigenfrequency $6\,\omega$. The laser beam passes underneath the masses.}
  \label{fig:PlanView}
\end{figure}

\section{Experimental Probe of Gravitational Dissipation}
\label{sec:Experiments}

\subsection{High Frequency Eotvos Experiment}
\label{sec:GeneralExperiment}
Here we discuss one possible experimental approach for testing the existence of dissipative gravity,
this approach roughly models the paradigm of the classic experiments of
Baron von Eotvos~\cite{eotvos1922contributions, EotVos:Redux} while incorporating the experience gained
over the past several decades of short scale gravity
experiments~\cite{Newman:CQG2001, Jens:Grav2005, Jens:Grav2007,  Geraci:2008hp, Rajalakshmi:2010cg}
The experiment consiststs of an active mechanical Attractor and a high sensitivity gravitational Responder.
They are constructed such that their oscillations cause tidal gravitational interactions.
By measuring the response of the Respnder, we can  look for the possible phase lag predicted
by a dissipative gravity model.

The outline of this Section goes as follows:
in Section~\ref{subsec:Hertz_sub_1:newtonian}, we estimate the magnitude of the gravitational
acceleration induced by our periodic attractor system.
in Section~\ref{subsec:Hertz_sub_2:signature}, we estimate the sensitivity limit of a mechanical
responder which is limied by Brownian thermal noise.
in Section~\ref{subsec:Hertz_sub_4:setup_noise_budget}, we calculate the minimum detectable
phase lag using the system we prescribe.

\begin{figure*}[ht]
\centering
\includegraphics[width=6in]{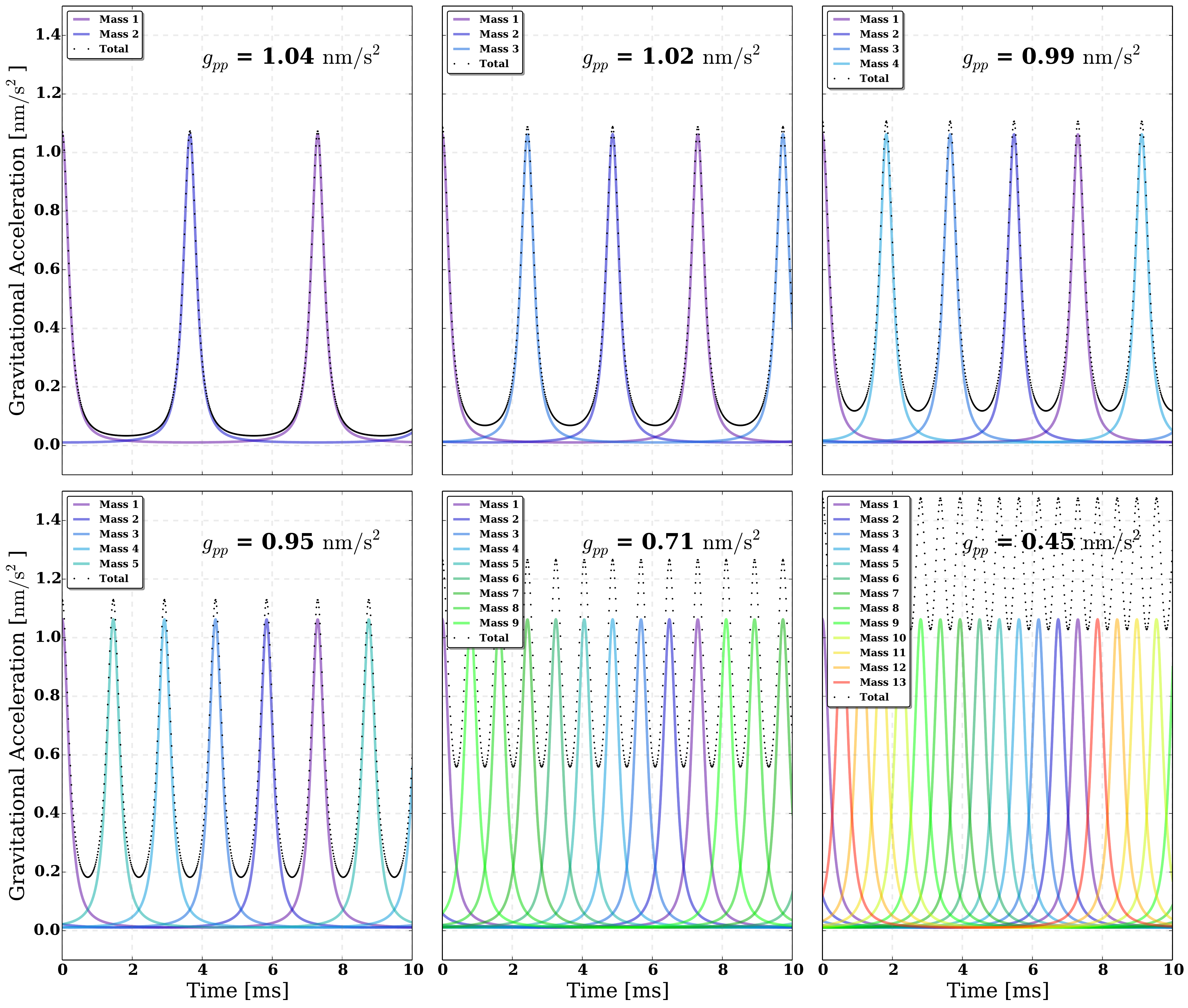}
  \caption{Differential acceleration (peak-peak) between the two ends of a Fabry-Perot cavity with a
  multi-mass periodic Attractor (cf.~Fig.~\ref{fig:PlanView}). As the number of
  masses is increased, the signal frequency increases, although the peak
  strain is correspondingly reduced. In all 6 plots, $m = 0.01\,kg$,
  $R = 0.095\,m$, and $d_1 = 0.12\,m$.}
  \label{fig:MultiBallRotor}
\end{figure*}

\subsubsection{Tidal gravitational field of the periodic attractor}
\label{subsec:Hertz_sub_1:newtonian}
Our aim is to probe the frequency dependence of gravitational forces and how they may differ, in the
non-relativistic limit, from Newtonian gravity. To this end we aim to use a system of moving masses that
create a periodic tidal gravitational attraction and will use a sensitive mechanical system to measure the
mechanical response to the gravitational force. An example of such a system is shown in Figure \ref{fig:PlanView}.

The scaling of the tidal gravitational acceleration of the periodic attractor system is approximated by
\begin{equation}
g_\mathrm{tidal} = G \rho L,
\end{equation}
where $G$ is Newton's constant, $\rho$ is the density of the material used in the
Attractor, and $L$ is a characteristc length scale of the attractor.

\subsubsection{Acceleration sensitivity of the responder}
\label{subsec:Hertz_sub_2:signature}
To maximize the signal amplitude, we propose to use a mechanical system with a high quality factor resonance
as our probe of the gravitational force. The motion of the mechanical system will be sensed using
interferometric metrology to maximize readout sensitivity. With such a narrow mechanical resonance,
and in addition we will integrate for long times to accumulate larger signal to noise ratio, for a given resonator
system the measurement is essentially single frequency.

A detailed noise analysis will be presented in Section \ref{subsec:Hertz_sub_4:setup_noise_budget}, and it will show that in most cases the dominant noise source is Brownian thermal fluctuations of the mechanical resonator.

Near the mechanical resonance, the force power spectral density of thermal fluctuations seen by the mechanical responder is given by
\begin{equation}
S_F^\mathrm{th} = 4mk_BT\frac{\omega_0}{Q},
\end{equation}
where $k_B$ is Boltzmann's constant, $m$ is the mass of the responder, $T$ is the temperature,
$\omega_0$ is the resonant frequency, and $Q$ is the mechanical quality factor.

Because we want to measure the sensitivity to gravitational acceleration, in units of acceleration power
spectral density, the thermal noise is
\begin{equation}
S_g^\mathrm{th} = 4k_BT\frac{\omega_0}{mQ}.
\end{equation}

\subsubsection{Sensitivity to phase lag}
\label{subsec:Hertz_sub_3:scaling}
The behavior of the mechanical resonance forces the Newtonian signal component to be in the
orthogonal quadrature with respect to the signal of the periodic attractor. And, ignoring systematic effects
which will be discussed in Section~\ref{sec:systematics}, any in-phase signal will be due to the
dissipative gravity effect. The magnitude of the signal in acceleration units is
\begin{equation}
g_\phi = \phi(\omega) g_\mathrm{tidal},
\end{equation}
and this is being compared to the RMS thermal acceleration of the oscillator due to thermal noise,
\begin{equation}
\sqrt{\langle g_\mathrm{th}^2 \rangle} = \sqrt{\frac{S_g^\mathrm{th}}{\tau}},
\label{eq:gnoiseandrms}
\end{equation}
where $\tau$ is the integration time, and we assume that $S_g$ is approximately constant near the
mechanical resonance. Therefore the minimum detectable dissipation phase angle detectable with an
SNR of unity after a time $\tau$ is
\begin{equation}
\phi(\omega_0) = \frac{1}{G\rho L}\sqrt{\frac{4k_BT\omega_0}{mQ\tau}}.
\end{equation}

Given the calculation shown in Figure \ref{fig:MultiBallRotor}, we can estimate that
$g_\mathrm{tidal} \approx 1\,\mathrm{nm}/\mathrm{s}^2$, and assuming
$\omega_0=2\pi\times 2\,\mathrm{kHz}$, $Q=10^8$, $T=1\,\mathrm{K}$,
$m=0.3\,\mathrm{kg}$, and $\tau=24\,\mathrm{hours}$, the minimim detectable phase
angle is $\phi=3.7 \times10^{-7}\,\mathrm{radians}$.

\begin{table}
    \begin{tabular}{lcccc}
    Parameter                  & Cantilever & 
    \begin{tabular}{c} Diluted \\ Cantilever \end{tabular} &  \begin{tabular}{c} Rigid \\ Cavity \end{tabular} & Pendula  \\ \hline \hline
 \begin{tabular}{c} Responder \\ Mass \end{tabular}             &   1\,mg &    1\,mg        &   300\,g     &   1\,kg           \\ \hline
    Responder Q                &    $10^6$  &     $10^{10}$         &      $10^8$  &  $10^8$           \\ \hline
     \begin{tabular}{c} Lowest \\ Eigenfrequency \end{tabular}      &   200\,Hz   &    40\,kHz          &   10\,kHz    &   2\,Hz           \\ \hline
    Temperature                &   100\,mK   &    100\,mK             &    1\,K      &   120\,K          \\ \hline
    Stored Power               &   10\,mW      &    60\,W             &    1\,kW     &   10\,W          \\ \hline
    \end{tabular}
        \caption{Physical parameters of proposed gravitational responders.}
    \label{tab:ExpParams}
\end{table}

\begin{figure*}[t]
\begin{subfigure}[\;Cantilever Cavity]{
\includegraphics[width=0.22\linewidth]{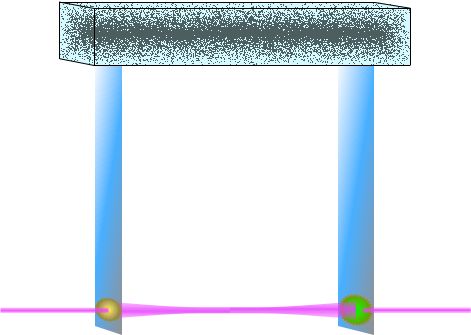}
    \label{fig:cavconf_a}
    }
      \end{subfigure}\hfill
\begin{subfigure}[\;Cantilever Cavity with Optical Spring Dilution]
{\includegraphics[width=0.22\linewidth]{CantileverCavity.png}
    \label{fig:cavconf_b}}
    \end{subfigure} \hfill
\begin{subfigure}[\;Massive Rigid Cavity]
{    \includegraphics[width=0.22\linewidth]{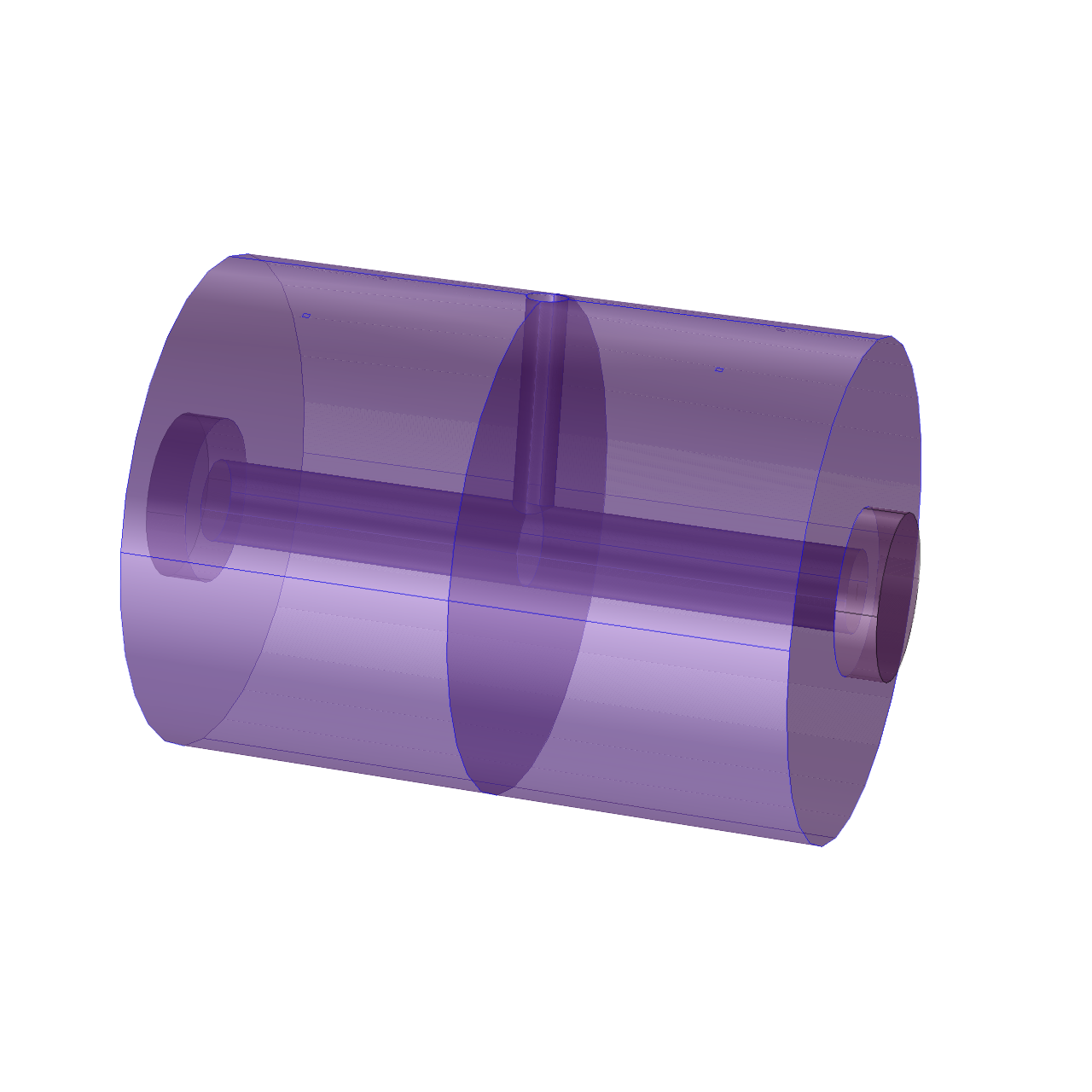}
    \label{fig:cavconf_c}}
    \end{subfigure} \hfill
\begin{subfigure}[\;Massive Pendula]
{    \includegraphics[width=0.22\linewidth]{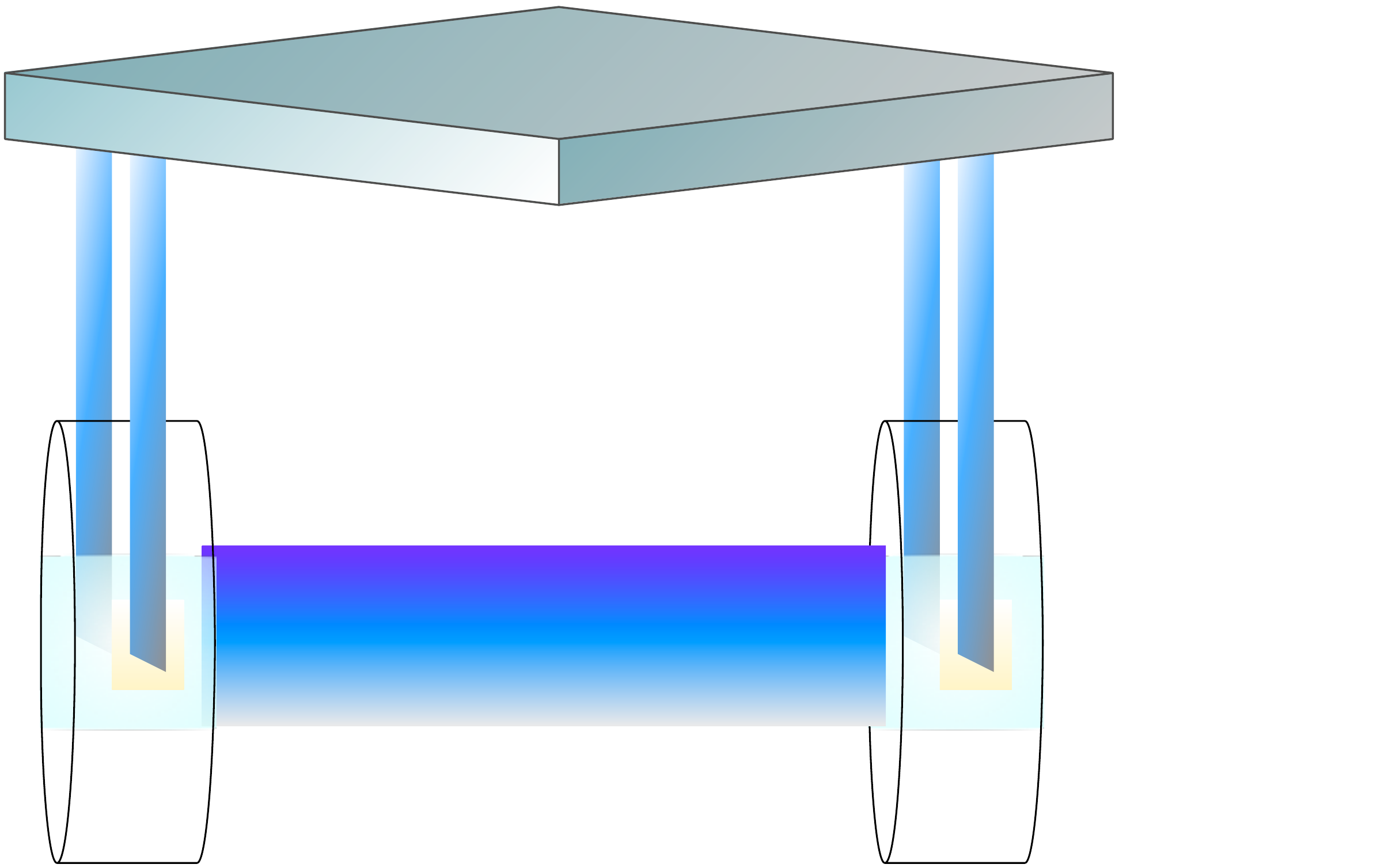}
    \label{fig:cavconf_d}}\end{subfigure}
  \caption{Schematic diagrams of four possible gravitational responders.}
  \label{fig:cavconf}
  
\end{figure*}

\subsection{Experimental Designs and Sensitivity Limits}
\label{subsec:Hertz_sub_4:setup_noise_budget}
The example given in Section\,\ref{subsec:Hertz_sub_1:newtonian} is illustrative
in showing the general order of magnitude of an expected signal. For a more
detailed analysis including a noise budget estimate, it is necessary to define
a specific resonator cavity geometry and readout scheme.

Here we describe a few potential Responder configurations that are common in
the gravitational-wave and opto-mechanics communities. All take the form
of a high finesse Fabry Perot cavity but vary in the configuration and
geometry of the cavity mirrors.

Schematic diagrams of the configurations are given in Figure\,\ref{fig:cavconf}.
Here we will briefly discuss the features of each configuration and
provide a noise budget estimate assuming feasible experimental
parameters. These parameters for each configuration are given
in Table\,\ref{tab:ExpParams}.

\begin{figure*}[t]
  \centering
    \includegraphics[width=6.75in]{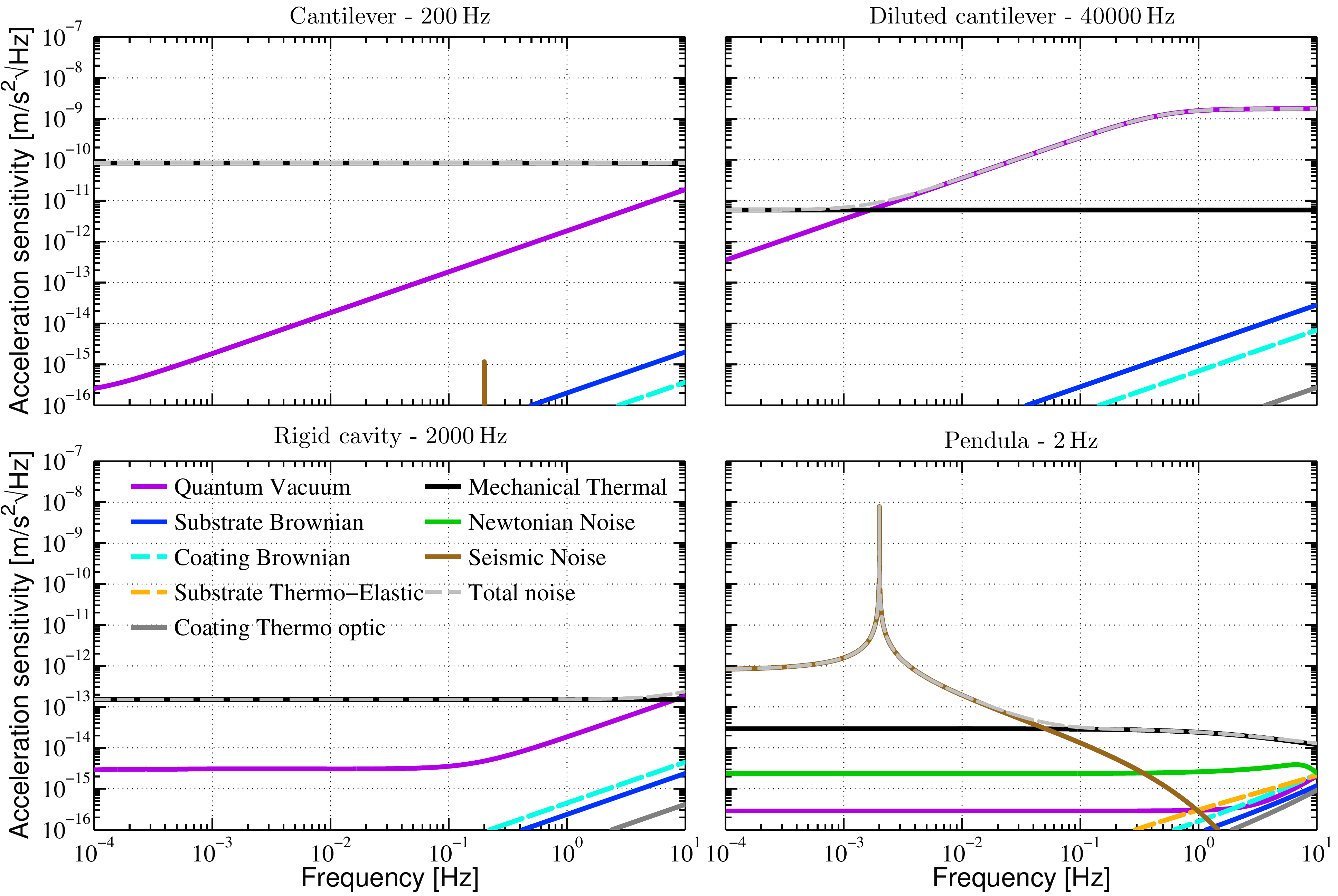}
    \caption{Shown here are the detailed noise budgets for the four proposed cavity configurations.
      The horizontal axis denotes positive sideband frequencies separated from the cavity mechanical
      resonance. In most cases the total noise is dominated by thermal noise of the mechanical oscillator.
      In the case of the suspended pendulum, the low frequency noise is dominated by seismic vibrations.}
   \label{fig:NB}
\end{figure*}

A pendulum suspended test mass configuration would be similar to the arm
cavities used in large scale gravitational-wave observatories such as
LIGO, however with reduced physical dimensions.

In this configuration, the sensing cavity is similar to the rigid kind used
for stabilization of lasers. A solid piece of single crystal silicon with
optically contacted mirrors is used to measure the gravitational accelerations.

In the past few decades, microcantilever force sensors have become widely
deployed as high sensitivity force and acceleration sensors~\cite{Boisen:2011hb},
especially in the biological agent detection fields and radiological. Microcantilevers have
also been used for sensitive fundamental physics
experiments\cite{Torres:2013ir},
such as
Casimir force experiments~\cite{2012PhRvB..85p5443C} and the searches for
fifth force and extra dimensions by
looking for short range deviations in the Newtonian $1/r^2$ law. They are recently also
candidates to measure quantum backaction
noise~\cite{SafaviNaeini:2013cr, Purdy:2013wf}.

\subsection{Components of the Noise Model}
Our noise model consists of several noise terms which limit the measurement of
the gravitational coupling of the periodic attractor to the responder. Each noise
term is explained below.

To characterize the Responder, in particular, we are interested in the sensitivity
to tidal gravitational accelerations. To
calculate the acceleration sensitivity, we first compute the displacement
noise contribution from all terms, then this is multiplied
by $|\chi^{-1}/m|$ to produce the acceleration noise. As described in
Section~\ref{subsec:Hertz_sub_3:scaling}, the sensitivity is maximized at the
mechanical resonance, and the resonance is very narrow, thus we make
logarithmic plots in Figure~\ref{fig:NB}, where the origin of the horizontal
axes are at the mechanical resonance frequencies.

\begin{figure*}[ht]
  \begin{center}
    \includegraphics[width=6.5in]{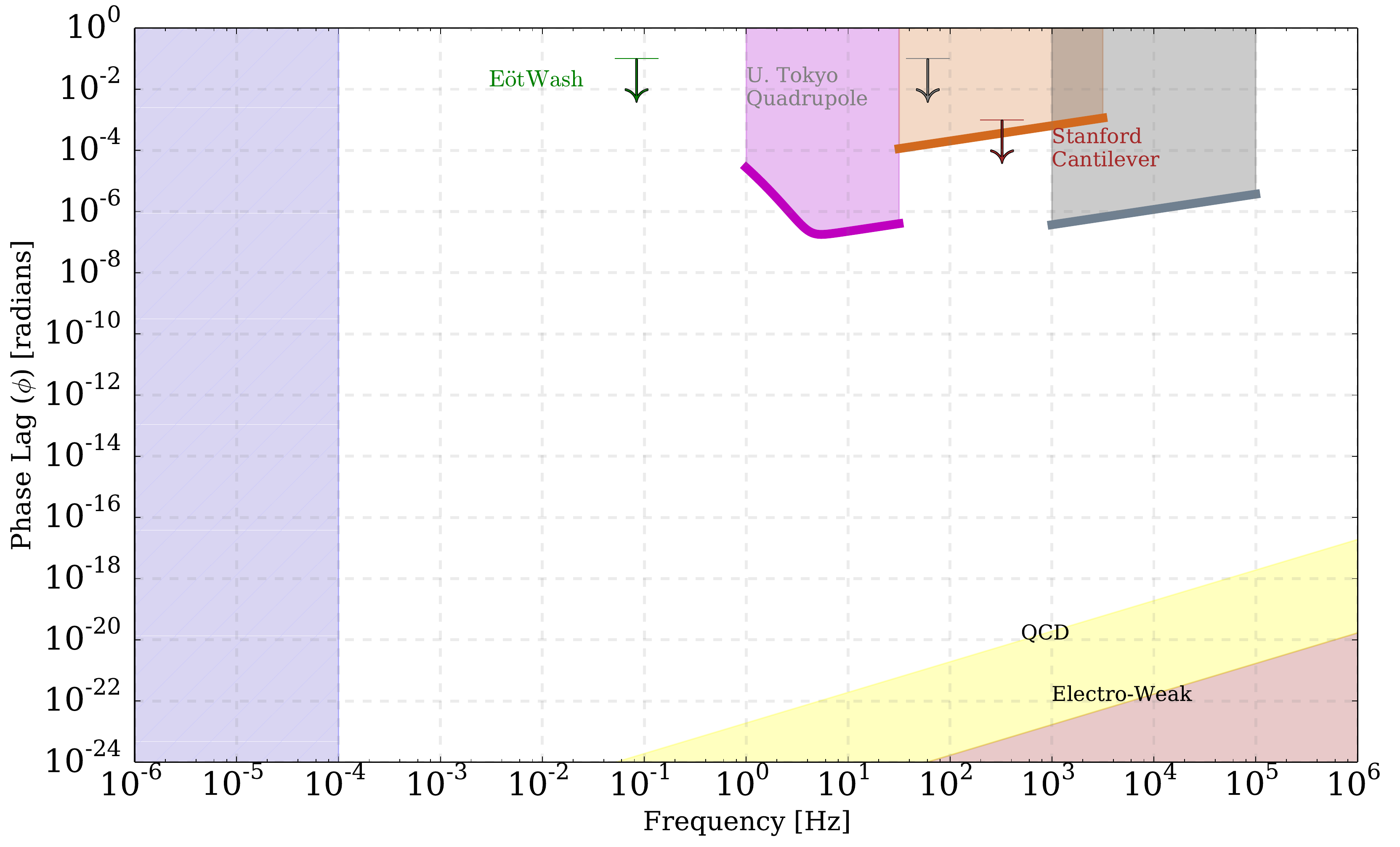}
    \caption{\textbf{Comparison of Experimental limits and Theoretical Predictions}.
    The arrows represent existing upper limits on the phase lag angle
    (cf.~Table~\ref{tab:constraints}). The shaded blue region is excluded due to pulsar observations.
    To give a sense of scale, the shaded regions at the bottom of the plot
    give estimates for $\phi$ assuming a dissipation of unity at the Electro-Weak and
    QCD length scales.
    The solid lines are expected upper limits from the experiments described in
    Sec.~\ref{subsec:Hertz_sub_4:setup_noise_budget}, where each point in the
    line represents an oscillator resonant at that frequency.
    The thick purple line (from 1-30 Hz) represents the massive pendulum cavity.
    The thick orange line (from 30-3000 Hz) represents the small cantilever cavity.
    The thick grey line (from 1-100 kHz) represents the rigid spacer cavity.}
   \label{fig:gravityexclusion}
 \end{center}
\end{figure*}

\subsubsection{Force Noises}
\paragraph{Seismic}
Vibrations of the laboratory due to external seismic fluctuations and nearby
vibrating machinery will produce motion of the support point of the Responder
cavity. This will be mostly rejected below the Responder's mechanical resonance
frequency. Near the resonance the rejection will be imperfect due to imperfect
matching of the pendular resonant frequencies (we assume a mismatch of
$\sim\,0.1\%$) of the two ends of the cavity.

\paragraph{Newtonian Gravitational Noise}
Fluctuations in the Newtonian gravitational potential due to local sources
(e.g. seismic, anthropogenic, air pressure, etc.) can limit the acceleration
sensitivity due to their spatial gradients across the ends of the Responder
cavity~\cite{NN:subtract2012, Teviet:2008}. As most of these fluctuations
are sourced by mechanical vibrations (and have a rather red power spectrum),
they are most serious for the massive suspended cavity and least serious for
the rigid, monolithic cavity (cf.~Fig.~\ref{fig:cavconf}).

\paragraph{Thermal Noise}
Several different kinds of thermal noise are relevant for these experiments.
Internal friction of the suspension support elements, cantilever flex joints, and
rigid cavity spacer result in Brownian forces which move the mirrors. The same
kind of internal friction in the mirror coatings also produces
Brownian~\cite{Ting:Brownian2012} strain fluctuations of the coating phase.
Thermodynamic temperature fluctuations~\cite{Mazo195957, BGV1999} can lead to
apparent optical cavity length fluctuations through a finite thermal
expansion coefficient of the mirror substrate (Substrate Thermo-Elastic)
and coatings, as well as through the temperature dependence of the index
of refraction of the coating materials~\cite{BGV2000, Levin20081941}. We refer
to the coherent expansion and refractive fluctuations in the coating as
`Coating Thermo-Optic'.

\paragraph{Quantum Vacuum}
Fluctuations of the vacuum electro-magnetic field limit a variety of
high precision measurements~\cite{Clerk:RMP2010, Yanbei:MQMreview}. In
the particular case of a Fabry-Perot interferometer, the fluctuations in
the phase quadrature limit the resolution of the optical phase measurement
and is usually referred to as 'shot noise'. The fluctuations in the amplitude
quadrature produce radiation pressure `shot noise' in the circulating
field within the Fabry-Perot cavity; this fluctuating pressure moves the cavity
mirrors. This Quantum Vacuum noise limits these experiments in different frequency
bands. If the thermal noise can be reduced, then squeezed light techniques
can be employed to further constrain the spacetime dissipation.

\subsection{Limits on complex phase angle for oscillators of different frequencies}
\label{subsec:AngleLimits}
The sensitivity to tidal accelerations given from our noise model can be combined with the
acceleration induced by the periodic attractor to give a final measurement limit of the
complex gravitational phase angle. Generalizing Equation \ref{eq:gnoiseandrms} for a
frequency varying $S_g$, the measurable limit on $\phi$ is

\begin{equation}
  \phi_\mathrm{lim} = \frac{1}{g_\mathrm{att}} \sqrt{\displaystyle \int_{\omega_0-1/2\tau_\mathrm{avg}}^{\omega_0+1/2\tau_\mathrm{avg}}S_g \frac{\mathrm{d}\omega}{2\pi}}
 \label{eq:AccSens}
\end{equation}
where $S_g$ is the \emph{power} spectral density of the acceleration noise of the receiver,
$\omega_0$ is the receiver resonance frequency, and $g_\mathrm{att}$ is the differential acceleration
caused by the periodic attractor. We estimate an acceleration on the order of
$1\,\mathrm{nm}/\mathrm{s}^2$ for $g_\mathrm{att}$, as illustrated in Figure~\ref{fig:MultiBallRotor}.
The limits for the various types of oscillator, for various possible resonant frequencies, after
24~hours of integration, are shown in Figure~\ref{fig:gravityexclusion}.

\subsection{Elimination of Systematic Effects}
\label{sec:systematics}
The sensitivity shown above in Figure~\ref{fig:gravityexclusion} includes only the
limits due to Gaussian random noise. In reality, experiments where the signal is so
weak are likely to be limited by systematic effects. In these searches for a
gravitational anomaly, we should be especially concerned about an anomalous signal
showing up due to imperfections in the experimental apparatus, rather than new physics.

There are several systematic effects including time delay, mechanical response
limits, mis-estimation of source oscillator phase, clock / timing errors in the
digitizer, timing error in the interferometric measurement of the source motion, etc.

We will estimate and subtract these effects in the following ways:
\begin{enumerate}
  \item Electro-magnetic pickup of the electrical driving signal in the readout
        electronics at $f_{\rm drive}$ will be rejected since it is not at the gravitational
        perturbation frequency ($f_{\rm signal}$). High frequency harmonics of the drive
        signal will be attenuated by usual grounding and shielding
        practices~\cite{morrison1998grounding}.

  \item Patch charges on the Attractor or Responder can produce a fluctuating electrostatic
        force at $f_{\rm motion}$. Both the Attractor and Responder can be coated with a slightly
        conductive coating to minimize charge gradients and the Attractor will be enclosed within
        a Faraday cage to ground these fields.

  \item Light scattered from the cavity mirrors can backscatter from the Attractor back into the
        cavity mode to produce a signal at $f_{\rm drive}$. To mitigate this, we will use super
        polished optics for low scatter and blacken the nearby surfaces. To calibrate the scatter
        signal we will apply a few Lambertian marks to the Attractor at irregular intervals.

\item In order to make a high precision estimate of $\phi$, one must know the mechanical admittance
         precisely. The phase angle of the admittance will be measured using non-gravitational forces
         (e.g. with electrostatic force actuators and radiation pressure on the laser beam). The difference
         between the electromagnetic force admittance and the gravitational admittance will ultimately
         limit the sensitivity of this search.
\end{enumerate}

\section{Conclusions}
\label{sec:conclusions}
The confrontation between quantum mechanics and general relativity points to a modification
of one or both of the theories. Previously, two communities have been tackling these problems: modification of gravity, and possible modifications of quantum mechanics.  In this work, we intend to bring up the connection between these two different approaches.  In order to do so, we have written down model theories that are both be viewed as modifications to GR and to QM.  Experimentally, these can plausibly be probed by precision measurement performed in the  laboratory scale. While it is
certain that the proposed experimental approaches are extremely challenging, they should be
able to, at the very least, place interesting upper limits on these types of alternative theories of
gravity and point us towards new physics.

\begin{acknowledgments}
We thank Rai Weiss, Eric K. Gustafson, Jan Harms, Nico Yunes, Yuri Levin, Poghos
Kazarian and John Preskill for illuminating discussions. We acknowledge funding provided by the Institute for Quantum
Information and Matter, an NSF Physics Frontiers Center with support of
the Gordon and Betty Moore Foundation.
HY, HM, and YC are supported by NSF Grant PHY-0601459, PHY-0653653, CAREER Grant PHY-0956189, and
the David and Barbara Groce Startup Fund at the California Institute of Technology.
RXA, NDSL and LP are supported by the National Science Foundation under grant PHY-0555406.
Research at Perimeter
Institute is supported through Industry Canada and by the Province of Ontario
through the Ministry of Research \& Innovation.
\end{acknowledgments}

\bibliography{GWreferences,references}

\begin{thebibliography}{80}%
\makeatletter
\providecommand \@ifxundefined [1]{%
 \@ifx{#1\undefined}
}%
\providecommand \@ifnum [1]{%
 \ifnum #1\expandafter \@firstoftwo
 \else \expandafter \@secondoftwo
 \fi
}%
\providecommand \@ifx [1]{%
 \ifx #1\expandafter \@firstoftwo
 \else \expandafter \@secondoftwo
 \fi
}%
\providecommand \natexlab [1]{#1}%
\providecommand \enquote  [1]{``#1''}%
\providecommand \bibnamefont  [1]{#1}%
\providecommand \bibfnamefont [1]{#1}%
\providecommand \citenamefont [1]{#1}%
\providecommand \href@noop [0]{\@secondoftwo}%
\providecommand \href [0]{\begingroup \@sanitize@url \@href}%
\providecommand \@href[1]{\@@startlink{#1}\@@href}%
\providecommand \@@href[1]{\endgroup#1\@@endlink}%
\providecommand \@sanitize@url [0]{\catcode `\\12\catcode `\$12\catcode
  `\&12\catcode `\#12\catcode `\^12\catcode `\_12\catcode `\%12\relax}%
\providecommand \@@startlink[1]{}%
\providecommand \@@endlink[0]{}%
\providecommand \url  [0]{\begingroup\@sanitize@url \@url }%
\providecommand \@url [1]{\endgroup\@href {#1}{\urlprefix }}%
\providecommand \urlprefix  [0]{URL }%
\providecommand \Eprint [0]{\href }%
\providecommand \doibase [0]{http://dx.doi.org/}%
\providecommand \selectlanguage [0]{\@gobble}%
\providecommand \bibinfo  [0]{\@secondoftwo}%
\providecommand \bibfield  [0]{\@secondoftwo}%
\providecommand \translation [1]{[#1]}%
\providecommand \BibitemOpen [0]{}%
\providecommand \bibitemStop [0]{}%
\providecommand \bibitemNoStop [0]{.\EOS\space}%
\providecommand \EOS [0]{\spacefactor3000\relax}%
\providecommand \BibitemShut  [1]{\csname bibitem#1\endcsname}%
\let\auto@bib@innerbib\@empty
%</preamble>
\bibitem [{\citenamefont {Will}(2006)}]{will2006confrontation}%
  \BibitemOpen
  \bibfield  {author} {\bibinfo {author} {\bibfnamefont {C.~M.}\ \bibnamefont
  {Will}},\ }\href@noop {} {\bibfield  {journal} {\bibinfo  {journal} {Living
  Rev. Relativity}\ }\textbf {\bibinfo {volume} {9}} (\bibinfo {year}
  {2006})}\BibitemShut {NoStop}%
\bibitem [{\citenamefont {Berti}\ \emph {et~al.}(2015)\citenamefont {Berti},
  \citenamefont {Barausse}, \citenamefont {Cardoso}, \citenamefont {Gualtieri},
  \citenamefont {Pani}, \citenamefont {Sperhake}, \citenamefont {Stein},
  \citenamefont {Wex}, \citenamefont {Yagi}, \citenamefont {Baker} \emph
  {et~al.}}]{berti2015testing}%
  \BibitemOpen
  \bibfield  {author} {\bibinfo {author} {\bibfnamefont {E.}~\bibnamefont
  {Berti}}, \bibinfo {author} {\bibfnamefont {E.}~\bibnamefont {Barausse}},
  \bibinfo {author} {\bibfnamefont {V.}~\bibnamefont {Cardoso}}, \bibinfo
  {author} {\bibfnamefont {L.}~\bibnamefont {Gualtieri}}, \bibinfo {author}
  {\bibfnamefont {P.}~\bibnamefont {Pani}}, \bibinfo {author} {\bibfnamefont
  {U.}~\bibnamefont {Sperhake}}, \bibinfo {author} {\bibfnamefont {L.~C.}\
  \bibnamefont {Stein}}, \bibinfo {author} {\bibfnamefont {N.}~\bibnamefont
  {Wex}}, \bibinfo {author} {\bibfnamefont {K.}~\bibnamefont {Yagi}}, \bibinfo
  {author} {\bibfnamefont {T.}~\bibnamefont {Baker}},  \emph {et~al.},\
  }\href@noop {} {\bibfield  {journal} {\bibinfo  {journal} {arXiv preprint
  arXiv:1501.07274}\ } (\bibinfo {year} {2015})}\BibitemShut {NoStop}%
\bibitem [{\citenamefont {Marshall}\ \emph {et~al.}(2003)\citenamefont
  {Marshall}, \citenamefont {Simon}, \citenamefont {Penrose},\ and\
  \citenamefont {Bouwmeester}}]{marshall2003towards}%
  \BibitemOpen
  \bibfield  {author} {\bibinfo {author} {\bibfnamefont {W.}~\bibnamefont
  {Marshall}}, \bibinfo {author} {\bibfnamefont {C.}~\bibnamefont {Simon}},
  \bibinfo {author} {\bibfnamefont {R.}~\bibnamefont {Penrose}}, \ and\
  \bibinfo {author} {\bibfnamefont {D.}~\bibnamefont {Bouwmeester}},\
  }\href@noop {} {\bibfield  {journal} {\bibinfo  {journal} {Physical Review
  Letters}\ }\textbf {\bibinfo {volume} {91}},\ \bibinfo {pages} {130401}
  (\bibinfo {year} {2003})}\BibitemShut {NoStop}%
\bibitem [{\citenamefont {Romero-Isart}\ \emph {et~al.}(2011)\citenamefont
  {Romero-Isart}, \citenamefont {Pflanzer}, \citenamefont {Blaser},
  \citenamefont {Kaltenbaek}, \citenamefont {Kiesel}, \citenamefont
  {Aspelmeyer},\ and\ \citenamefont {Cirac}}]{romero2011large}%
  \BibitemOpen
  \bibfield  {author} {\bibinfo {author} {\bibfnamefont {O.}~\bibnamefont
  {Romero-Isart}}, \bibinfo {author} {\bibfnamefont {A.}~\bibnamefont
  {Pflanzer}}, \bibinfo {author} {\bibfnamefont {F.}~\bibnamefont {Blaser}},
  \bibinfo {author} {\bibfnamefont {R.}~\bibnamefont {Kaltenbaek}}, \bibinfo
  {author} {\bibfnamefont {N.}~\bibnamefont {Kiesel}}, \bibinfo {author}
  {\bibfnamefont {M.}~\bibnamefont {Aspelmeyer}}, \ and\ \bibinfo {author}
  {\bibfnamefont {J.}~\bibnamefont {Cirac}},\ }\href@noop {} {\bibfield
  {journal} {\bibinfo  {journal} {Physical Review Letters}\ }\textbf {\bibinfo
  {volume} {107}},\ \bibinfo {pages} {020405} (\bibinfo {year}
  {2011})}\BibitemShut {NoStop}%
\bibitem [{\citenamefont {Pikovski}\ \emph {et~al.}(2012)\citenamefont
  {Pikovski}, \citenamefont {Vanner}, \citenamefont {Aspelmeyer}, \citenamefont
  {Kim},\ and\ \citenamefont {Brukner}}]{pikovski2012probing}%
  \BibitemOpen
  \bibfield  {author} {\bibinfo {author} {\bibfnamefont {I.}~\bibnamefont
  {Pikovski}}, \bibinfo {author} {\bibfnamefont {M.~R.}\ \bibnamefont
  {Vanner}}, \bibinfo {author} {\bibfnamefont {M.}~\bibnamefont {Aspelmeyer}},
  \bibinfo {author} {\bibfnamefont {M.}~\bibnamefont {Kim}}, \ and\ \bibinfo
  {author} {\bibfnamefont {{\v{C}}.}~\bibnamefont {Brukner}},\ }\href@noop {}
  {\bibfield  {journal} {\bibinfo  {journal} {Nature Physics}\ }\textbf
  {\bibinfo {volume} {8}},\ \bibinfo {pages} {393} (\bibinfo {year}
  {2012})}\BibitemShut {NoStop}%
\bibitem [{\citenamefont {Kaltenbaek}\ \emph {et~al.}(2015)\citenamefont
  {Kaltenbaek}, \citenamefont {Arndt}, \citenamefont {Aspelmeyer},
  \citenamefont {Barker}, \citenamefont {Bassi}, \citenamefont {Bateman},
  \citenamefont {Bongs}, \citenamefont {Bose}, \citenamefont {Braxmaier},
  \citenamefont {Brukner} \emph {et~al.}}]{kaltenbaek2015macroscopic}%
  \BibitemOpen
  \bibfield  {author} {\bibinfo {author} {\bibfnamefont {R.}~\bibnamefont
  {Kaltenbaek}}, \bibinfo {author} {\bibfnamefont {M.}~\bibnamefont {Arndt}},
  \bibinfo {author} {\bibfnamefont {M.}~\bibnamefont {Aspelmeyer}}, \bibinfo
  {author} {\bibfnamefont {P.~F.}\ \bibnamefont {Barker}}, \bibinfo {author}
  {\bibfnamefont {A.}~\bibnamefont {Bassi}}, \bibinfo {author} {\bibfnamefont
  {J.}~\bibnamefont {Bateman}}, \bibinfo {author} {\bibfnamefont
  {K.}~\bibnamefont {Bongs}}, \bibinfo {author} {\bibfnamefont
  {S.}~\bibnamefont {Bose}}, \bibinfo {author} {\bibfnamefont {C.}~\bibnamefont
  {Braxmaier}}, \bibinfo {author} {\bibfnamefont {{\v{C}}.}~\bibnamefont
  {Brukner}},  \emph {et~al.},\ }\href@noop {} {\bibfield  {journal} {\bibinfo
  {journal} {arXiv preprint arXiv:1503.02640}\ } (\bibinfo {year}
  {2015})}\BibitemShut {NoStop}%
\bibitem [{\citenamefont {Kaltenbaek}\ \emph {et~al.}(2012)\citenamefont
  {Kaltenbaek}, \citenamefont {Hechenblaikner}, \citenamefont {Kiesel},
  \citenamefont {Romero-Isart}, \citenamefont {Schwab}, \citenamefont
  {Johann},\ and\ \citenamefont {Aspelmeyer}}]{kaltenbaek2012macroscopic}%
  \BibitemOpen
  \bibfield  {author} {\bibinfo {author} {\bibfnamefont {R.}~\bibnamefont
  {Kaltenbaek}}, \bibinfo {author} {\bibfnamefont {G.}~\bibnamefont
  {Hechenblaikner}}, \bibinfo {author} {\bibfnamefont {N.}~\bibnamefont
  {Kiesel}}, \bibinfo {author} {\bibfnamefont {O.}~\bibnamefont
  {Romero-Isart}}, \bibinfo {author} {\bibfnamefont {K.~C.}\ \bibnamefont
  {Schwab}}, \bibinfo {author} {\bibfnamefont {U.}~\bibnamefont {Johann}}, \
  and\ \bibinfo {author} {\bibfnamefont {M.}~\bibnamefont {Aspelmeyer}},\
  }\href@noop {} {\bibfield  {journal} {\bibinfo  {journal} {Experimental
  Astronomy}\ }\textbf {\bibinfo {volume} {34}},\ \bibinfo {pages} {123}
  (\bibinfo {year} {2012})}\BibitemShut {NoStop}%
\bibitem [{\citenamefont {Nimmrichter}\ and\ \citenamefont
  {Hornberger}(2013)}]{nimmrichter2013macroscopicity}%
  \BibitemOpen
  \bibfield  {author} {\bibinfo {author} {\bibfnamefont {S.}~\bibnamefont
  {Nimmrichter}}\ and\ \bibinfo {author} {\bibfnamefont {K.}~\bibnamefont
  {Hornberger}},\ }\href@noop {} {\bibfield  {journal} {\bibinfo  {journal}
  {Physical review letters}\ }\textbf {\bibinfo {volume} {110}},\ \bibinfo
  {pages} {160403} (\bibinfo {year} {2013})}\BibitemShut {NoStop}%
\bibitem [{\citenamefont {Yang}\ \emph {et~al.}(2013)\citenamefont {Yang},
  \citenamefont {Miao}, \citenamefont {Lee}, \citenamefont {Helou},\ and\
  \citenamefont {Chen}}]{yang2013macroscopic}%
  \BibitemOpen
  \bibfield  {author} {\bibinfo {author} {\bibfnamefont {H.}~\bibnamefont
  {Yang}}, \bibinfo {author} {\bibfnamefont {H.}~\bibnamefont {Miao}}, \bibinfo
  {author} {\bibfnamefont {D.-S.}\ \bibnamefont {Lee}}, \bibinfo {author}
  {\bibfnamefont {B.}~\bibnamefont {Helou}}, \ and\ \bibinfo {author}
  {\bibfnamefont {Y.}~\bibnamefont {Chen}},\ }\href@noop {} {\bibfield
  {journal} {\bibinfo  {journal} {Physical review letters}\ }\textbf {\bibinfo
  {volume} {110}},\ \bibinfo {pages} {170401} (\bibinfo {year}
  {2013})}\BibitemShut {NoStop}%
\bibitem [{\citenamefont {Nimmrichter}\ \emph {et~al.}(2014)\citenamefont
  {Nimmrichter}, \citenamefont {Hornberger},\ and\ \citenamefont
  {Hammerer}}]{nimmrichter2014optomechanical}%
  \BibitemOpen
  \bibfield  {author} {\bibinfo {author} {\bibfnamefont {S.}~\bibnamefont
  {Nimmrichter}}, \bibinfo {author} {\bibfnamefont {K.}~\bibnamefont
  {Hornberger}}, \ and\ \bibinfo {author} {\bibfnamefont {K.}~\bibnamefont
  {Hammerer}},\ }\href@noop {} {\bibfield  {journal} {\bibinfo  {journal}
  {Physical review letters}\ }\textbf {\bibinfo {volume} {113}},\ \bibinfo
  {pages} {020405} (\bibinfo {year} {2014})}\BibitemShut {NoStop}%
\bibitem [{\citenamefont {Di\'osi}(2015)}]{diosi2015testing}%
  \BibitemOpen
  \bibfield  {author} {\bibinfo {author} {\bibfnamefont {L.}~\bibnamefont
  {Di\'osi}},\ }\href {\doibase 10.1103/PhysRevLett.114.050403} {\bibfield
  {journal} {\bibinfo  {journal} {Phys. Rev. Lett.}\ }\textbf {\bibinfo
  {volume} {114}},\ \bibinfo {pages} {050403} (\bibinfo {year}
  {2015})}\BibitemShut {NoStop}%
\bibitem [{\citenamefont {Chen}(2013{\natexlab{a}})}]{chen2013macroscopic}%
  \BibitemOpen
  \bibfield  {author} {\bibinfo {author} {\bibfnamefont {Y.}~\bibnamefont
  {Chen}},\ }\href@noop {} {\bibfield  {journal} {\bibinfo  {journal} {Journal
  of Physics B: Atomic, Molecular and Optical Physics}\ }\textbf {\bibinfo
  {volume} {46}},\ \bibinfo {pages} {104001} (\bibinfo {year}
  {2013}{\natexlab{a}})}\BibitemShut {NoStop}%
\bibitem [{\citenamefont {Kafri}\ \emph {et~al.}(2015)\citenamefont {Kafri},
  \citenamefont {Milburn},\ and\ \citenamefont {Taylor}}]{kafri2015bounds}%
  \BibitemOpen
  \bibfield  {author} {\bibinfo {author} {\bibfnamefont {D.}~\bibnamefont
  {Kafri}}, \bibinfo {author} {\bibfnamefont {G.}~\bibnamefont {Milburn}}, \
  and\ \bibinfo {author} {\bibfnamefont {J.}~\bibnamefont {Taylor}},\
  }\href@noop {} {\bibfield  {journal} {\bibinfo  {journal} {New Journal of
  Physics}\ }\textbf {\bibinfo {volume} {17}},\ \bibinfo {pages} {015006}
  (\bibinfo {year} {2015})}\BibitemShut {NoStop}%
\bibitem [{\citenamefont {Pfister}\ \emph {et~al.}(2015)\citenamefont
  {Pfister}, \citenamefont {Kaniewski}, \citenamefont {Tomamichel},
  \citenamefont {Mantri}, \citenamefont {Schmucker}, \citenamefont {McMahon},
  \citenamefont {Milburn},\ and\ \citenamefont
  {Wehner}}]{pfister2015understanding}%
  \BibitemOpen
  \bibfield  {author} {\bibinfo {author} {\bibfnamefont {C.}~\bibnamefont
  {Pfister}}, \bibinfo {author} {\bibfnamefont {J.}~\bibnamefont {Kaniewski}},
  \bibinfo {author} {\bibfnamefont {M.}~\bibnamefont {Tomamichel}}, \bibinfo
  {author} {\bibfnamefont {A.}~\bibnamefont {Mantri}}, \bibinfo {author}
  {\bibfnamefont {R.}~\bibnamefont {Schmucker}}, \bibinfo {author}
  {\bibfnamefont {N.}~\bibnamefont {McMahon}}, \bibinfo {author} {\bibfnamefont
  {G.}~\bibnamefont {Milburn}}, \ and\ \bibinfo {author} {\bibfnamefont
  {S.}~\bibnamefont {Wehner}},\ }\href@noop {} {\bibfield  {journal} {\bibinfo
  {journal} {arXiv preprint arXiv:1503.00577}\ } (\bibinfo {year}
  {2015})}\BibitemShut {NoStop}%
\bibitem [{\citenamefont {Hu}(2012)}]{hu2012emergence}%
  \BibitemOpen
  \bibfield  {author} {\bibinfo {author} {\bibfnamefont {B.-L.}\ \bibnamefont
  {Hu}},\ }in\ \href@noop {} {\emph {\bibinfo {booktitle} {Journal of Physics:
  Conference Series}}},\ Vol.\ \bibinfo {volume} {361}\ (\bibinfo
  {organization} {IOP Publishing},\ \bibinfo {year} {2012})\ p.\ \bibinfo
  {pages} {012003}\BibitemShut {NoStop}%
\bibitem [{\citenamefont {Sakharov}(1968)}]{Sakharov:1967pk}%
  \BibitemOpen
  \bibfield  {author} {\bibinfo {author} {\bibfnamefont {A.}~\bibnamefont
  {Sakharov}},\ }\href@noop {} {\bibfield  {journal} {\bibinfo  {journal}
  {Sov.Phys.Dokl.}\ }\textbf {\bibinfo {volume} {12}},\ \bibinfo {pages} {1040}
  (\bibinfo {year} {1968})}\BibitemShut {NoStop}%
%%CITATION = SPHDA,12,1040;%%
\bibitem [{\citenamefont {Hu}(1996)}]{hu1996general}%
  \BibitemOpen
  \bibfield  {author} {\bibinfo {author} {\bibfnamefont {B.}~\bibnamefont
  {Hu}},\ }\href@noop {} {\bibfield  {journal} {\bibinfo  {journal} {arXiv
  preprint gr-qc/9607070}\ } (\bibinfo {year} {1996})}\BibitemShut {NoStop}%
\bibitem [{\citenamefont {Bardeen}\ \emph {et~al.}(1973)\citenamefont
  {Bardeen}, \citenamefont {Carter},\ and\ \citenamefont
  {Hawking}}]{Bardeen:1973gs}%
  \BibitemOpen
  \bibfield  {author} {\bibinfo {author} {\bibfnamefont {J.~M.}\ \bibnamefont
  {Bardeen}}, \bibinfo {author} {\bibfnamefont {B.}~\bibnamefont {Carter}}, \
  and\ \bibinfo {author} {\bibfnamefont {S.}~\bibnamefont {Hawking}},\ }\href
  {\doibase 10.1007/BF01645742} {\bibfield  {journal} {\bibinfo  {journal}
  {Commun.Math.Phys.}\ }\textbf {\bibinfo {volume} {31}},\ \bibinfo {pages}
  {161} (\bibinfo {year} {1973})}\BibitemShut {NoStop}%
%%CITATION = CMPHA,31,161;%%
\bibitem [{\citenamefont {Hawking}(1975)}]{Hawking:1974sw}%
  \BibitemOpen
  \bibfield  {author} {\bibinfo {author} {\bibfnamefont {S.}~\bibnamefont
  {Hawking}},\ }\href {\doibase 10.1007/BF02345020, 10.1007/BF02345020}
  {\bibfield  {journal} {\bibinfo  {journal} {Commun.Math.Phys.}\ }\textbf
  {\bibinfo {volume} {43}},\ \bibinfo {pages} {199} (\bibinfo {year}
  {1975})}\BibitemShut {NoStop}%
%%CITATION = CMPHA,43,199;%%
\bibitem [{\citenamefont {Jacobson}(1995)}]{Jacobson:1995ab}%
  \BibitemOpen
  \bibfield  {author} {\bibinfo {author} {\bibfnamefont {T.}~\bibnamefont
  {Jacobson}},\ }\href {\doibase 10.1103/PhysRevLett.75.1260} {\bibfield
  {journal} {\bibinfo  {journal} {Phys.Rev.Lett.}\ }\textbf {\bibinfo {volume}
  {75}},\ \bibinfo {pages} {1260} (\bibinfo {year} {1995})},\ \Eprint
  {http://arxiv.org/abs/gr-qc/9504004} {arXiv:gr-qc/9504004 [gr-qc]}
  \BibitemShut {NoStop}%
%%CITATION = GR-QC/9504004;%%
\bibitem [{\citenamefont {Verlinde}(2011)}]{Verlinde:2010hp}%
  \BibitemOpen
  \bibfield  {author} {\bibinfo {author} {\bibfnamefont {E.~P.}\ \bibnamefont
  {Verlinde}},\ }\href {\doibase 10.1007/JHEP04(2011)029} {\bibfield  {journal}
  {\bibinfo  {journal} {JHEP}\ }\textbf {\bibinfo {volume} {1104}},\ \bibinfo
  {pages} {029} (\bibinfo {year} {2011})},\ \Eprint
  {http://arxiv.org/abs/1001.0785} {arXiv:1001.0785 [hep-th]} \BibitemShut
  {NoStop}%
%%CITATION = ARXIV:1001.0785;%%
\bibitem [{\citenamefont {Smolin}(2010)}]{Smolin:2010kk}%
  \BibitemOpen
  \bibfield  {author} {\bibinfo {author} {\bibfnamefont {L.}~\bibnamefont
  {Smolin}},\ }\href@noop {} {\  (\bibinfo {year} {2010})},\ \Eprint
  {http://arxiv.org/abs/1001.3668} {arXiv:1001.3668 [gr-qc]} \BibitemShut
  {NoStop}%
%%CITATION = ARXIV:1001.3668;%%
\bibitem [{\citenamefont {Reynaud}\ and\ \citenamefont
  {Jaekel}(2008)}]{Reynaud:2008ts}%
  \BibitemOpen
  \bibfield  {author} {\bibinfo {author} {\bibfnamefont {S.}~\bibnamefont
  {Reynaud}}\ and\ \bibinfo {author} {\bibfnamefont {M.-T.}\ \bibnamefont
  {Jaekel}},\ }\href@noop {} {\bibfield  {journal} {\bibinfo  {journal}
  {arXiv.org}\ } (\bibinfo {year} {2008})}\BibitemShut {NoStop}%
\bibitem [{\citenamefont {Turyshev}(2010)}]{Tury:2010}%
  \BibitemOpen
  \bibfield  {author} {\bibinfo {author} {\bibfnamefont {S.~G.}\ \bibnamefont
  {Turyshev}},\ }in\ \href@noop {} {\emph {\bibinfo {booktitle} {GRAVITATIONAL
  PHYSICS: TESTING GRAVITY FROM SUBMILLIMETER TO COSMIC: Proceedings of the
  VIII Mexican School on Gravitation and Mathematical Physics. AIP Conference
  Proceedings}}}\ (\bibinfo {organization} {Jet Propulsion Laboratory,
  California Institute of Technology, 4800 Oak Grove Drive, Pasadena, CA
  91109-0899, USA},\ \bibinfo {year} {2010})\ pp.\ \bibinfo {pages}
  {3--26}\BibitemShut {NoStop}%
\bibitem [{\citenamefont {Taylor}(1994)}]{Taylor:1994RMP}%
  \BibitemOpen
  \bibfield  {author} {\bibinfo {author} {\bibfnamefont {J.~H.~J.}\
  \bibnamefont {Taylor}},\ }\href@noop {} {\bibfield  {journal} {\bibinfo
  {journal} {Reviews of Modern Physics}\ }\textbf {\bibinfo {volume} {66}},\
  \bibinfo {pages} {711} (\bibinfo {year} {1994})}\BibitemShut {NoStop}%
\bibitem [{\citenamefont {Kramer}\ \emph {et~al.}(2006)\citenamefont {Kramer},
  \citenamefont {Stairs}, \citenamefont {Manchester}, \citenamefont
  {McLaughlin}, \citenamefont {Lyne}, \citenamefont {Ferdman}, \citenamefont
  {Burgay}, \citenamefont {Lorimer}, \citenamefont {Possenti}, \citenamefont
  {D'Amico}, \citenamefont {Sarkissian}, \citenamefont {Joshi}, \citenamefont
  {Freire},\ and\ \citenamefont {Camilo}}]{Kramer:2006cl}%
  \BibitemOpen
  \bibfield  {author} {\bibinfo {author} {\bibfnamefont {M.}~\bibnamefont
  {Kramer}}, \bibinfo {author} {\bibfnamefont {I.~H.}\ \bibnamefont {Stairs}},
  \bibinfo {author} {\bibfnamefont {R.~N.}\ \bibnamefont {Manchester}},
  \bibinfo {author} {\bibfnamefont {M.~A.}\ \bibnamefont {McLaughlin}},
  \bibinfo {author} {\bibfnamefont {A.~G.}\ \bibnamefont {Lyne}}, \bibinfo
  {author} {\bibfnamefont {R.~D.}\ \bibnamefont {Ferdman}}, \bibinfo {author}
  {\bibfnamefont {M.}~\bibnamefont {Burgay}}, \bibinfo {author} {\bibfnamefont
  {D.~R.}\ \bibnamefont {Lorimer}}, \bibinfo {author} {\bibfnamefont
  {A.}~\bibnamefont {Possenti}}, \bibinfo {author} {\bibfnamefont
  {N.}~\bibnamefont {D'Amico}}, \bibinfo {author} {\bibfnamefont
  {J.}~\bibnamefont {Sarkissian}}, \bibinfo {author} {\bibfnamefont {B.~C.}\
  \bibnamefont {Joshi}}, \bibinfo {author} {\bibfnamefont {P.~C.~C.}\
  \bibnamefont {Freire}}, \ and\ \bibinfo {author} {\bibfnamefont
  {F.}~\bibnamefont {Camilo}},\ }\href@noop {} {\bibfield  {journal} {\bibinfo
  {journal} {Annalen der Physik}\ }\textbf {\bibinfo {volume} {15}},\ \bibinfo
  {pages} {34} (\bibinfo {year} {2006})}\BibitemShut {NoStop}%
\bibitem [{\citenamefont {Beane}(1997)}]{Beane:1997kl}%
  \BibitemOpen
  \bibfield  {author} {\bibinfo {author} {\bibfnamefont {S.~R.}\ \bibnamefont
  {Beane}},\ }\href@noop {} {\bibfield  {journal} {\bibinfo  {journal} {General
  Relativity and Gravitation}\ }\textbf {\bibinfo {volume} {29}},\ \bibinfo
  {pages} {945} (\bibinfo {year} {1997})}\BibitemShut {NoStop}%
\bibitem [{\citenamefont {Randall}(1999)}]{Randall:1999eq}%
  \BibitemOpen
  \bibfield  {author} {\bibinfo {author} {\bibfnamefont {L.}~\bibnamefont
  {Randall}},\ }\href@noop {} {\bibfield  {journal} {\bibinfo  {journal}
  {Physical Review Letters}\ }\textbf {\bibinfo {volume} {83}},\ \bibinfo
  {pages} {3370} (\bibinfo {year} {1999})}\BibitemShut {NoStop}%
\bibitem [{\citenamefont {Arkani~Hamed}\ \emph {et~al.}(1998)\citenamefont
  {Arkani~Hamed}, \citenamefont {Dimopoulos},\ and\ \citenamefont
  {Dvali}}]{ArkaniHamed:1998gb}%
  \BibitemOpen
  \bibfield  {author} {\bibinfo {author} {\bibfnamefont {N.}~\bibnamefont
  {Arkani~Hamed}}, \bibinfo {author} {\bibfnamefont {S.}~\bibnamefont
  {Dimopoulos}}, \ and\ \bibinfo {author} {\bibfnamefont {G.}~\bibnamefont
  {Dvali}},\ }\href@noop {} {\bibfield  {journal} {\bibinfo  {journal} {Physics
  Letters B}\ }\textbf {\bibinfo {volume} {429}},\ \bibinfo {pages} {263}
  (\bibinfo {year} {1998})}\BibitemShut {NoStop}%
\bibitem [{\citenamefont {{Gundlach}}(2005)}]{Jens:Grav2005}%
  \BibitemOpen
  \bibfield  {author} {\bibinfo {author} {\bibfnamefont {J.~H.}\ \bibnamefont
  {{Gundlach}}},\ }\href {\doibase 10.1088/1367-2630/7/1/205} {\bibfield
  {journal} {\bibinfo  {journal} {New Journal of Physics}\ }\textbf {\bibinfo
  {volume} {7}},\ \bibinfo {pages} {205} (\bibinfo {year} {2005})}\BibitemShut
  {NoStop}%
\bibitem [{\citenamefont {Cocconi}\ and\ \citenamefont
  {Salpeter}(1960)}]{PhysRevLett.4.176}%
  \BibitemOpen
  \bibfield  {author} {\bibinfo {author} {\bibfnamefont {G.}~\bibnamefont
  {Cocconi}}\ and\ \bibinfo {author} {\bibfnamefont {E.~E.}\ \bibnamefont
  {Salpeter}},\ }\href {\doibase 10.1103/PhysRevLett.4.176} {\bibfield
  {journal} {\bibinfo  {journal} {Phys. Rev. Lett.}\ }\textbf {\bibinfo
  {volume} {4}},\ \bibinfo {pages} {176} (\bibinfo {year} {1960})}\BibitemShut
  {NoStop}%
\bibitem [{\citenamefont {Drever}(1961)}]{Drever:1961cl}%
  \BibitemOpen
  \bibfield  {author} {\bibinfo {author} {\bibfnamefont {R.~W.~P.}\
  \bibnamefont {Drever}},\ }\href@noop {} {\bibfield  {journal} {\bibinfo
  {journal} {Philosophical Magazine}\ }\textbf {\bibinfo {volume} {6}},\
  \bibinfo {pages} {683} (\bibinfo {year} {1961})}\BibitemShut {NoStop}%
\bibitem [{\citenamefont {Tasson}(2014)}]{Tasson:2014dv}%
  \BibitemOpen
  \bibfield  {author} {\bibinfo {author} {\bibfnamefont {J.~D.}\ \bibnamefont
  {Tasson}},\ }\href@noop {} {\bibfield  {journal} {\bibinfo  {journal} {Rep.
  Prog. Phys}\ }\textbf {\bibinfo {volume} {77}},\ \bibinfo {pages} {062901}
  (\bibinfo {year} {2014})}\BibitemShut {NoStop}%
\bibitem [{\citenamefont {Mattingly}(2005)}]{lrr-2005-5}%
  \BibitemOpen
  \bibfield  {author} {\bibinfo {author} {\bibfnamefont {D.}~\bibnamefont
  {Mattingly}},\ }\href {\doibase 10.12942/lrr-2005-5} {\bibfield  {journal}
  {\bibinfo  {journal} {Living Reviews in Relativity}\ }\textbf {\bibinfo
  {volume} {8}} (\bibinfo {year} {2005}),\ 10.12942/lrr-2005-5}\BibitemShut
  {NoStop}%
\bibitem [{\citenamefont {Michimura}\ \emph {et~al.}(2013)\citenamefont
  {Michimura}, \citenamefont {Mewes}, \citenamefont {Matsumoto}, \citenamefont
  {Aso},\ and\ \citenamefont {Ando}}]{Yuta:PRD13}%
  \BibitemOpen
  \bibfield  {author} {\bibinfo {author} {\bibfnamefont {Y.}~\bibnamefont
  {Michimura}}, \bibinfo {author} {\bibfnamefont {M.}~\bibnamefont {Mewes}},
  \bibinfo {author} {\bibfnamefont {N.}~\bibnamefont {Matsumoto}}, \bibinfo
  {author} {\bibfnamefont {Y.}~\bibnamefont {Aso}}, \ and\ \bibinfo {author}
  {\bibfnamefont {M.}~\bibnamefont {Ando}},\ }\href@noop {} {\bibfield
  {journal} {\bibinfo  {journal} {arXiv.org}\ ,\ \bibinfo {pages} {111101}}
  (\bibinfo {year} {2013})},\ \Eprint {http://arxiv.org/abs/1310.1952v2}
  {1310.1952v2} \BibitemShut {NoStop}%
\bibitem [{\citenamefont {Herrmann}\ \emph {et~al.}(2005)\citenamefont
  {Herrmann}, \citenamefont {Senger}, \citenamefont {Kovalchuk}, \citenamefont
  {M{\"u}ller},\ and\ \citenamefont {Peters}}]{Herrmann:2005tp}%
  \BibitemOpen
  \bibfield  {author} {\bibinfo {author} {\bibfnamefont {S.}~\bibnamefont
  {Herrmann}}, \bibinfo {author} {\bibfnamefont {A.}~\bibnamefont {Senger}},
  \bibinfo {author} {\bibfnamefont {E.}~\bibnamefont {Kovalchuk}}, \bibinfo
  {author} {\bibfnamefont {H.}~\bibnamefont {M{\"u}ller}}, \ and\ \bibinfo
  {author} {\bibfnamefont {A.}~\bibnamefont {Peters}},\ }\href@noop {}
  {\bibfield  {journal} {\bibinfo  {journal} {Physical Review Letters}\
  }\textbf {\bibinfo {volume} {95}},\ \bibinfo {pages} {150401} (\bibinfo
  {year} {2005})}\BibitemShut {NoStop}%
\bibitem [{\citenamefont {Hehl}\ and\ \citenamefont
  {Mashhoon}(2009)}]{hehl2009nonlocal}%
  \BibitemOpen
  \bibfield  {author} {\bibinfo {author} {\bibfnamefont {F.~W.}\ \bibnamefont
  {Hehl}}\ and\ \bibinfo {author} {\bibfnamefont {B.}~\bibnamefont
  {Mashhoon}},\ }\href@noop {} {\bibfield  {journal} {\bibinfo  {journal}
  {Physics Letters B}\ }\textbf {\bibinfo {volume} {673}},\ \bibinfo {pages}
  {279} (\bibinfo {year} {2009})}\BibitemShut {NoStop}%
\bibitem [{\citenamefont {Mashhoon}(2014)}]{mashhoon2014nonlocal}%
  \BibitemOpen
  \bibfield  {author} {\bibinfo {author} {\bibfnamefont {B.}~\bibnamefont
  {Mashhoon}},\ }\href@noop {} {\bibfield  {journal} {\bibinfo  {journal}
  {Physical Review D}\ }\textbf {\bibinfo {volume} {90}},\ \bibinfo {pages}
  {124031} (\bibinfo {year} {2014})}\BibitemShut {NoStop}%
\bibitem [{\citenamefont {Diosi}(1984)}]{diosi1984gravitation}%
  \BibitemOpen
  \bibfield  {author} {\bibinfo {author} {\bibfnamefont {L.}~\bibnamefont
  {Diosi}},\ }\href@noop {} {\bibfield  {journal} {\bibinfo  {journal} {Physics
  letters A}\ }\textbf {\bibinfo {volume} {105}},\ \bibinfo {pages} {199}
  (\bibinfo {year} {1984})}\BibitemShut {NoStop}%
\bibitem [{\citenamefont {Penrose}(1998)}]{penrose1998quantum}%
  \BibitemOpen
  \bibfield  {author} {\bibinfo {author} {\bibfnamefont {R.}~\bibnamefont
  {Penrose}},\ }\href@noop {} {\bibfield  {journal} {\bibinfo  {journal}
  {PHILOSOPHICAL TRANSACTIONS-ROYAL SOCIETY OF LONDON SERIES A MATHEMATICAL
  PHYSICAL AND ENGINEERING SCIENCES}\ ,\ \bibinfo {pages} {1927}} (\bibinfo
  {year} {1998})}\BibitemShut {NoStop}%
\bibitem [{\citenamefont {Ghirardi}\ \emph {et~al.}(1986)\citenamefont
  {Ghirardi}, \citenamefont {Rimini},\ and\ \citenamefont
  {Weber}}]{ghirardi1986unified}%
  \BibitemOpen
  \bibfield  {author} {\bibinfo {author} {\bibfnamefont {G.~C.}\ \bibnamefont
  {Ghirardi}}, \bibinfo {author} {\bibfnamefont {A.}~\bibnamefont {Rimini}}, \
  and\ \bibinfo {author} {\bibfnamefont {T.}~\bibnamefont {Weber}},\
  }\href@noop {} {\bibfield  {journal} {\bibinfo  {journal} {Physical Review
  D}\ }\textbf {\bibinfo {volume} {34}},\ \bibinfo {pages} {470} (\bibinfo
  {year} {1986})}\BibitemShut {NoStop}%
\bibitem [{\citenamefont {Ghirardi}\ \emph {et~al.}(1990)\citenamefont
  {Ghirardi}, \citenamefont {Pearle},\ and\ \citenamefont
  {Rimini}}]{ghirardi1990markov}%
  \BibitemOpen
  \bibfield  {author} {\bibinfo {author} {\bibfnamefont {G.~C.}\ \bibnamefont
  {Ghirardi}}, \bibinfo {author} {\bibfnamefont {P.}~\bibnamefont {Pearle}}, \
  and\ \bibinfo {author} {\bibfnamefont {A.}~\bibnamefont {Rimini}},\
  }\href@noop {} {\bibfield  {journal} {\bibinfo  {journal} {Physical Review
  A}\ }\textbf {\bibinfo {volume} {42}},\ \bibinfo {pages} {78} (\bibinfo
  {year} {1990})}\BibitemShut {NoStop}%
\bibitem [{\citenamefont {Di{\'o}si}(2014)}]{diosi2014gravity}%
  \BibitemOpen
  \bibfield  {author} {\bibinfo {author} {\bibfnamefont {L.}~\bibnamefont
  {Di{\'o}si}},\ }\href@noop {} {\bibfield  {journal} {\bibinfo  {journal}
  {Foundations of Physics}\ }\textbf {\bibinfo {volume} {44}},\ \bibinfo
  {pages} {483} (\bibinfo {year} {2014})}\BibitemShut {NoStop}%
\bibitem [{\citenamefont {Di\'osi}(2014)}]{Diosi:2014a}%
  \BibitemOpen
  \bibfield  {author} {\bibinfo {author} {\bibfnamefont {L.}~\bibnamefont
  {Di\'osi}},\ }\href@noop {} {\bibfield  {journal} {\bibinfo  {journal} {EPJ
  Web of Conferences}\ }\textbf {\bibinfo {volume} {78}},\ \bibinfo {pages}
  {02001} (\bibinfo {year} {2014})}\BibitemShut {NoStop}%
\bibitem [{\citenamefont {{Hong}}\ \emph {et~al.}(2013)\citenamefont {{Hong}},
  \citenamefont {{Yang}}, \citenamefont {{Gustafson}}, \citenamefont
  {{Adhikari}},\ and\ \citenamefont {{Chen}}}]{Ting:Brownian2012}%
  \BibitemOpen
  \bibfield  {author} {\bibinfo {author} {\bibfnamefont {T.}~\bibnamefont
  {{Hong}}}, \bibinfo {author} {\bibfnamefont {H.}~\bibnamefont {{Yang}}},
  \bibinfo {author} {\bibfnamefont {E.~K.}\ \bibnamefont {{Gustafson}}},
  \bibinfo {author} {\bibfnamefont {R.~X.}\ \bibnamefont {{Adhikari}}}, \ and\
  \bibinfo {author} {\bibfnamefont {Y.}~\bibnamefont {{Chen}}},\ }\href
  {\doibase 10.1103/PhysRevD.87.082001} {\bibfield  {journal} {\bibinfo
  {journal} {Physical Review D}\ }\textbf {\bibinfo {volume} {87}},\ \bibinfo
  {eid} {082001} (\bibinfo {year} {2013})},\ \Eprint
  {http://arxiv.org/abs/1207.6145} {arXiv:1207.6145 [gr-qc]} \BibitemShut
  {NoStop}%
\bibitem [{\citenamefont {Jacobson}(2007)}]{Jacobson2007}%
  \BibitemOpen
  \bibfield  {author} {\bibinfo {author} {\bibfnamefont {T.}~\bibnamefont
  {Jacobson}},\ }\href@noop {} {\bibfield  {journal} {\bibinfo  {journal}
  {PoSQG-Ph}\ ,\ \bibinfo {pages} {20}} (\bibinfo {year} {2007})},\ \Eprint
  {http://arxiv.org/abs/gr-qc/08011547} {arXiv:gr-qc/08011547 [gr-qc]}
  \BibitemShut {NoStop}%
\bibitem [{\citenamefont {P.~Nicolini}\ and\ \citenamefont
  {Spallucci}(2006)}]{Nicolini:2005vd}%
  \BibitemOpen
  \bibfield  {author} {\bibinfo {author} {\bibfnamefont {A.~S.}\ \bibnamefont
  {P.~Nicolini}}\ and\ \bibinfo {author} {\bibfnamefont {E.}~\bibnamefont
  {Spallucci}},\ }\href@noop {} {\bibfield  {journal} {\bibinfo  {journal}
  {Phys.~Lett.~B}\ }\textbf {\bibinfo {volume} {632}},\ \bibinfo {pages} {547}
  (\bibinfo {year} {2006})}\BibitemShut {NoStop}%
\bibitem [{\citenamefont {L.~Modesto}\ and\ \citenamefont
  {Nicolini}(2011)}]{Modesto:2010uh}%
  \BibitemOpen
  \bibfield  {author} {\bibinfo {author} {\bibfnamefont {J.~W.~M.}\
  \bibnamefont {L.~Modesto}}\ and\ \bibinfo {author} {\bibfnamefont
  {P.}~\bibnamefont {Nicolini}},\ }\href@noop {} {\bibfield  {journal}
  {\bibinfo  {journal} {Phys.~Lett.~B}\ }\textbf {\bibinfo {volume} {695}},\
  \bibinfo {pages} {695} (\bibinfo {year} {2011})}\BibitemShut {NoStop}%
\bibitem [{\citenamefont {{Kramer}}\ \emph {et~al.}(2006)\citenamefont
  {{Kramer}}, \citenamefont {{Stairs}}, \citenamefont {{Manchester}},
  \citenamefont {{McLaughlin}}, \citenamefont {{Lyne}}, \citenamefont
  {{Ferdman}}, \citenamefont {{Burgay}}, \citenamefont {{Lorimer}},
  \citenamefont {{Possenti}}, \citenamefont {{D'Amico}}, \citenamefont
  {{Sarkissian}}, \citenamefont {{Hobbs}}, \citenamefont {{Reynolds}},
  \citenamefont {{Freire}},\ and\ \citenamefont
  {{Camilo}}}]{2006Sci...314...97K}%
  \BibitemOpen
  \bibfield  {author} {\bibinfo {author} {\bibfnamefont {M.}~\bibnamefont
  {{Kramer}}}, \bibinfo {author} {\bibfnamefont {I.~H.}\ \bibnamefont
  {{Stairs}}}, \bibinfo {author} {\bibfnamefont {R.~N.}\ \bibnamefont
  {{Manchester}}}, \bibinfo {author} {\bibfnamefont {M.~A.}\ \bibnamefont
  {{McLaughlin}}}, \bibinfo {author} {\bibfnamefont {A.~G.}\ \bibnamefont
  {{Lyne}}}, \bibinfo {author} {\bibfnamefont {R.~D.}\ \bibnamefont
  {{Ferdman}}}, \bibinfo {author} {\bibfnamefont {M.}~\bibnamefont {{Burgay}}},
  \bibinfo {author} {\bibfnamefont {D.~R.}\ \bibnamefont {{Lorimer}}}, \bibinfo
  {author} {\bibfnamefont {A.}~\bibnamefont {{Possenti}}}, \bibinfo {author}
  {\bibfnamefont {N.}~\bibnamefont {{D'Amico}}}, \bibinfo {author}
  {\bibfnamefont {J.~M.}\ \bibnamefont {{Sarkissian}}}, \bibinfo {author}
  {\bibfnamefont {G.~B.}\ \bibnamefont {{Hobbs}}}, \bibinfo {author}
  {\bibfnamefont {J.~E.}\ \bibnamefont {{Reynolds}}}, \bibinfo {author}
  {\bibfnamefont {P.~C.~C.}\ \bibnamefont {{Freire}}}, \ and\ \bibinfo {author}
  {\bibfnamefont {F.}~\bibnamefont {{Camilo}}},\ }\href {\doibase
  10.1126/science.1132305} {\bibfield  {journal} {\bibinfo  {journal}
  {Science}\ }\textbf {\bibinfo {volume} {314}},\ \bibinfo {pages} {97}
  (\bibinfo {year} {2006})},\ \Eprint {http://arxiv.org/abs/astro-ph/0609417}
  {astro-ph/0609417} \BibitemShut {NoStop}%
\bibitem [{\citenamefont {{Stairs}}\ \emph {et~al.}(2002)\citenamefont
  {{Stairs}}, \citenamefont {{Thorsett}}, \citenamefont {{Taylor}},\ and\
  \citenamefont {{Wolszczan}}}]{2002ApJ...581..501S}%
  \BibitemOpen
  \bibfield  {author} {\bibinfo {author} {\bibfnamefont {I.~H.}\ \bibnamefont
  {{Stairs}}}, \bibinfo {author} {\bibfnamefont {S.~E.}\ \bibnamefont
  {{Thorsett}}}, \bibinfo {author} {\bibfnamefont {J.~H.}\ \bibnamefont
  {{Taylor}}}, \ and\ \bibinfo {author} {\bibfnamefont {A.}~\bibnamefont
  {{Wolszczan}}},\ }\href {\doibase 10.1086/344157} {\bibfield  {journal}
  {\bibinfo  {journal} {\apj}\ }\textbf {\bibinfo {volume} {581}},\ \bibinfo
  {pages} {501} (\bibinfo {year} {2002})},\ \Eprint
  {http://arxiv.org/abs/astro-ph/0208357} {astro-ph/0208357} \BibitemShut
  {NoStop}%
\bibitem [{\citenamefont {{Weisberg}}\ \emph {et~al.}(2010)\citenamefont
  {{Weisberg}}, \citenamefont {{Nice}},\ and\ \citenamefont
  {{Taylor}}}]{2010ApJ...722.1030W}%
  \BibitemOpen
  \bibfield  {author} {\bibinfo {author} {\bibfnamefont {J.~M.}\ \bibnamefont
  {{Weisberg}}}, \bibinfo {author} {\bibfnamefont {D.~J.}\ \bibnamefont
  {{Nice}}}, \ and\ \bibinfo {author} {\bibfnamefont {J.~H.}\ \bibnamefont
  {{Taylor}}},\ }\href {\doibase 10.1088/0004-637X/722/2/1030} {\bibfield
  {journal} {\bibinfo  {journal} {\apj}\ }\textbf {\bibinfo {volume} {722}},\
  \bibinfo {pages} {1030} (\bibinfo {year} {2010})},\ \Eprint
  {http://arxiv.org/abs/1011.0718} {arXiv:1011.0718 [astro-ph.GA]} \BibitemShut
  {NoStop}%
\bibitem [{\citenamefont {{Jacoby}}\ \emph {et~al.}(2006)\citenamefont
  {{Jacoby}}, \citenamefont {{Cameron}}, \citenamefont {{Jenet}}, \citenamefont
  {{Anderson}}, \citenamefont {{Murty}},\ and\ \citenamefont
  {{Kulkarni}}}]{2006ApJ...644L.113J}%
  \BibitemOpen
  \bibfield  {author} {\bibinfo {author} {\bibfnamefont {B.~A.}\ \bibnamefont
  {{Jacoby}}}, \bibinfo {author} {\bibfnamefont {P.~B.}\ \bibnamefont
  {{Cameron}}}, \bibinfo {author} {\bibfnamefont {F.~A.}\ \bibnamefont
  {{Jenet}}}, \bibinfo {author} {\bibfnamefont {S.~B.}\ \bibnamefont
  {{Anderson}}}, \bibinfo {author} {\bibfnamefont {R.~N.}\ \bibnamefont
  {{Murty}}}, \ and\ \bibinfo {author} {\bibfnamefont {S.~R.}\ \bibnamefont
  {{Kulkarni}}},\ }\href {\doibase 10.1086/505742} {\bibfield  {journal}
  {\bibinfo  {journal} {\apjl}\ }\textbf {\bibinfo {volume} {644}},\ \bibinfo
  {pages} {L113} (\bibinfo {year} {2006})},\ \Eprint
  {http://arxiv.org/abs/astro-ph/0605375} {astro-ph/0605375} \BibitemShut
  {NoStop}%
\bibitem [{\citenamefont {Stephenson}(1997)}]{Stepehson1997}%
  \BibitemOpen
  \bibfield  {author} {\bibinfo {author} {\bibfnamefont {F.~R.}\ \bibnamefont
  {Stephenson}},\ }\href@noop {} {\emph {\bibinfo {title} {Historical eclipse
  and Earth's rotation}}}\ (\bibinfo  {publisher} {Cambridge university
  press},\ \bibinfo {year} {1997})\BibitemShut {NoStop}%
\bibitem [{\citenamefont {Kopeikin}(2001)}]{Kopeikin:2001}%
  \BibitemOpen
  \bibfield  {author} {\bibinfo {author} {\bibfnamefont {S.~M.}\ \bibnamefont
  {Kopeikin}},\ }\href@noop {} {\bibfield  {journal} {\bibinfo  {journal}
  {Astrophys. J.}\ }\textbf {\bibinfo {volume} {556}},\ \bibinfo {pages} {L1}
  (\bibinfo {year} {2001})}\BibitemShut {NoStop}%
\bibitem [{\citenamefont {Fomalont}\ and\ \citenamefont
  {Kopeikin}(2003)}]{Fomalont:2003}%
  \BibitemOpen
  \bibfield  {author} {\bibinfo {author} {\bibfnamefont {E.~B.}\ \bibnamefont
  {Fomalont}}\ and\ \bibinfo {author} {\bibfnamefont {S.~M.}\ \bibnamefont
  {Kopeikin}},\ }\href@noop {} {\bibfield  {journal} {\bibinfo  {journal}
  {Astrophys. J.}\ }\textbf {\bibinfo {volume} {598}},\ \bibinfo {pages} {704}
  (\bibinfo {year} {2003})}\BibitemShut {NoStop}%
\bibitem [{\citenamefont {Shapiro}(1966)}]{Shapiro:1966cp}%
  \BibitemOpen
  \bibfield  {author} {\bibinfo {author} {\bibfnamefont {I.}~\bibnamefont
  {Shapiro}},\ }\href@noop {} {\bibfield  {journal} {\bibinfo  {journal}
  {Physical Review}\ }\textbf {\bibinfo {volume} {141}},\ \bibinfo {pages}
  {1219} (\bibinfo {year} {1966})}\BibitemShut {NoStop}%
\bibitem [{\citenamefont {{Kapner}}\ \emph {et~al.}(2007)\citenamefont
  {{Kapner}}, \citenamefont {{Cook}}, \citenamefont {{Adelberger}},
  \citenamefont {{Gundlach}}, \citenamefont {{Heckel}}, \citenamefont
  {{Hoyle}},\ and\ \citenamefont {{Swanson}}}]{Jens:Grav2007}%
  \BibitemOpen
  \bibfield  {author} {\bibinfo {author} {\bibfnamefont {D.~J.}\ \bibnamefont
  {{Kapner}}}, \bibinfo {author} {\bibfnamefont {T.~S.}\ \bibnamefont
  {{Cook}}}, \bibinfo {author} {\bibfnamefont {E.~G.}\ \bibnamefont
  {{Adelberger}}}, \bibinfo {author} {\bibfnamefont {J.~H.}\ \bibnamefont
  {{Gundlach}}}, \bibinfo {author} {\bibfnamefont {B.~R.}\ \bibnamefont
  {{Heckel}}}, \bibinfo {author} {\bibfnamefont {C.~D.}\ \bibnamefont
  {{Hoyle}}}, \ and\ \bibinfo {author} {\bibfnamefont {H.~E.}\ \bibnamefont
  {{Swanson}}},\ }\href {\doibase 10.1103/PhysRevLett.98.021101} {\bibfield
  {journal} {\bibinfo  {journal} {Physical Review Letters}\ }\textbf {\bibinfo
  {volume} {98}},\ \bibinfo {eid} {021101} (\bibinfo {year} {2007})},\ \Eprint
  {http://arxiv.org/abs/hep-ph/0611184} {hep-ph/0611184} \BibitemShut {NoStop}%
\bibitem [{\citenamefont {Hirakawa}\ \emph {et~al.}(1980)\citenamefont
  {Hirakawa}, \citenamefont {Tsubono},\ and\ \citenamefont
  {Oide}}]{Hirakawa:1980fh}%
  \BibitemOpen
  \bibfield  {author} {\bibinfo {author} {\bibfnamefont {H.}~\bibnamefont
  {Hirakawa}}, \bibinfo {author} {\bibfnamefont {K.}~\bibnamefont {Tsubono}}, \
  and\ \bibinfo {author} {\bibfnamefont {K.}~\bibnamefont {Oide}},\ }\href@noop
  {} {\bibfield  {journal} {\bibinfo  {journal} {Nature}\ }\textbf {\bibinfo
  {volume} {283}},\ \bibinfo {pages} {184} (\bibinfo {year}
  {1980})}\BibitemShut {NoStop}%
\bibitem [{\citenamefont {Geraci}\ \emph {et~al.}(2008)\citenamefont {Geraci},
  \citenamefont {Smullin}, \citenamefont {Weld}, \citenamefont {Chiaverini},\
  and\ \citenamefont {Kapitulnik}}]{Geraci:2008hp}%
  \BibitemOpen
  \bibfield  {author} {\bibinfo {author} {\bibfnamefont {A.}~\bibnamefont
  {Geraci}}, \bibinfo {author} {\bibfnamefont {S.}~\bibnamefont {Smullin}},
  \bibinfo {author} {\bibfnamefont {D.}~\bibnamefont {Weld}}, \bibinfo {author}
  {\bibfnamefont {J.}~\bibnamefont {Chiaverini}}, \ and\ \bibinfo {author}
  {\bibfnamefont {A.}~\bibnamefont {Kapitulnik}},\ }\href@noop {} {\bibfield
  {journal} {\bibinfo  {journal} {Physical Review D}\ }\textbf {\bibinfo
  {volume} {78}},\ \bibinfo {pages} {022002} (\bibinfo {year}
  {2008})}\BibitemShut {NoStop}%
\bibitem [{\citenamefont {Weld}\ \emph {et~al.}(2008)\citenamefont {Weld},
  \citenamefont {Xia}, \citenamefont {Cabrera},\ and\ \citenamefont
  {Kapitulnik}}]{Weld:Gas2008}%
  \BibitemOpen
  \bibfield  {author} {\bibinfo {author} {\bibfnamefont {D.~M.}\ \bibnamefont
  {Weld}}, \bibinfo {author} {\bibfnamefont {J.}~\bibnamefont {Xia}}, \bibinfo
  {author} {\bibfnamefont {B.}~\bibnamefont {Cabrera}}, \ and\ \bibinfo
  {author} {\bibfnamefont {A.}~\bibnamefont {Kapitulnik}},\ }\href@noop {}
  {\bibfield  {journal} {\bibinfo  {journal} {Physical Review D}\ }\textbf
  {\bibinfo {volume} {77}},\ \bibinfo {pages} {62006} (\bibinfo {year}
  {2008})}\BibitemShut {NoStop}%
\bibitem [{\citenamefont {Hamilton}\ and\ \citenamefont
  {Crescimanno}(2008)}]{Hamilton2008}%
  \BibitemOpen
  \bibfield  {author} {\bibinfo {author} {\bibfnamefont {B.}~\bibnamefont
  {Hamilton}}\ and\ \bibinfo {author} {\bibfnamefont {M.}~\bibnamefont
  {Crescimanno}},\ }\href@noop {} {\bibfield  {journal} {\bibinfo  {journal}
  {Journal of Physics A: Mathematical and Theoretical}\ }\textbf {\bibinfo
  {volume} {41}},\ \bibinfo {pages} {235205} (\bibinfo {year}
  {2008})}\BibitemShut {NoStop}%
\bibitem [{\citenamefont {Demorest}\ \emph {et~al.}(2010)\citenamefont
  {Demorest}, \citenamefont {Pennucci}, \citenamefont {Ransom}, \citenamefont
  {Roberts},\ and\ \citenamefont {Hessels}}]{Demorest:2010bf}%
  \BibitemOpen
  \bibfield  {author} {\bibinfo {author} {\bibfnamefont {P.~B.}\ \bibnamefont
  {Demorest}}, \bibinfo {author} {\bibfnamefont {T.}~\bibnamefont {Pennucci}},
  \bibinfo {author} {\bibfnamefont {S.~M.}\ \bibnamefont {Ransom}}, \bibinfo
  {author} {\bibfnamefont {M.~S.~E.}\ \bibnamefont {Roberts}}, \ and\ \bibinfo
  {author} {\bibfnamefont {J.~W.~T.}\ \bibnamefont {Hessels}},\ }\href@noop {}
  {\bibfield  {journal} {\bibinfo  {journal} {Nature}\ }\textbf {\bibinfo
  {volume} {467}},\ \bibinfo {pages} {1081} (\bibinfo {year}
  {2010})}\BibitemShut {NoStop}%
\bibitem [{\citenamefont {Eotvos}\ \emph {et~al.}(1922)\citenamefont {Eotvos},
  \citenamefont {Pekar},\ and\ \citenamefont
  {Fekete}}]{eotvos1922contributions}%
  \BibitemOpen
  \bibfield  {author} {\bibinfo {author} {\bibfnamefont {R.~V.}\ \bibnamefont
  {Eotvos}}, \bibinfo {author} {\bibfnamefont {D.}~\bibnamefont {Pekar}}, \
  and\ \bibinfo {author} {\bibfnamefont {E.}~\bibnamefont {Fekete}},\
  }\href@noop {} {\bibfield  {journal} {\bibinfo  {journal} {Annalen der
  Physik}\ }\textbf {\bibinfo {volume} {68}},\ \bibinfo {pages} {11} (\bibinfo
  {year} {1922})}\BibitemShut {NoStop}%
\bibitem [{\citenamefont {{Fischbach}}\ \emph {et~al.}(1986)\citenamefont
  {{Fischbach}}, \citenamefont {{Sudarsky}}, \citenamefont {{Szafer}},
  \citenamefont {{Talmadge}},\ and\ \citenamefont {{Aronson}}}]{EotVos:Redux}%
  \BibitemOpen
  \bibfield  {author} {\bibinfo {author} {\bibfnamefont {E.}~\bibnamefont
  {{Fischbach}}}, \bibinfo {author} {\bibfnamefont {D.}~\bibnamefont
  {{Sudarsky}}}, \bibinfo {author} {\bibfnamefont {A.}~\bibnamefont
  {{Szafer}}}, \bibinfo {author} {\bibfnamefont {C.}~\bibnamefont
  {{Talmadge}}}, \ and\ \bibinfo {author} {\bibfnamefont {S.~H.}\ \bibnamefont
  {{Aronson}}},\ }\href {\doibase 10.1103/PhysRevLett.56.3} {\bibfield
  {journal} {\bibinfo  {journal} {Physical Review Letters}\ }\textbf {\bibinfo
  {volume} {56}},\ \bibinfo {pages} {3} (\bibinfo {year} {1986})}\BibitemShut
  {NoStop}%
\bibitem [{\citenamefont {{Newman}}(2001)}]{Newman:CQG2001}%
  \BibitemOpen
  \bibfield  {author} {\bibinfo {author} {\bibfnamefont {R.}~\bibnamefont
  {{Newman}}},\ }\href {\doibase 10.1088/0264-9381/18/13/303} {\bibfield
  {journal} {\bibinfo  {journal} {Classical and Quantum Gravity}\ }\textbf
  {\bibinfo {volume} {18}},\ \bibinfo {pages} {2407} (\bibinfo {year}
  {2001})}\BibitemShut {NoStop}%
\bibitem [{\citenamefont {Rajalakshmi}\ and\ \citenamefont
  {Unnikrishnan}(2010)}]{Rajalakshmi:2010cg}%
  \BibitemOpen
  \bibfield  {author} {\bibinfo {author} {\bibfnamefont {G.}~\bibnamefont
  {Rajalakshmi}}\ and\ \bibinfo {author} {\bibfnamefont {C.~S.}\ \bibnamefont
  {Unnikrishnan}},\ }\href@noop {} {\bibfield  {journal} {\bibinfo  {journal}
  {Classical and Quantum Gravity}\ }\textbf {\bibinfo {volume} {27}},\ \bibinfo
  {pages} {215007} (\bibinfo {year} {2010})}\BibitemShut {NoStop}%
\bibitem [{\citenamefont {Boisen}\ \emph {et~al.}(2011)\citenamefont {Boisen},
  \citenamefont {Dohn}, \citenamefont {Keller}, \citenamefont {Schmid},\ and\
  \citenamefont {Tenje}}]{Boisen:2011hb}%
  \BibitemOpen
  \bibfield  {author} {\bibinfo {author} {\bibfnamefont {A.}~\bibnamefont
  {Boisen}}, \bibinfo {author} {\bibfnamefont {S.}~\bibnamefont {Dohn}},
  \bibinfo {author} {\bibfnamefont {S.~S.}\ \bibnamefont {Keller}}, \bibinfo
  {author} {\bibfnamefont {S.}~\bibnamefont {Schmid}}, \ and\ \bibinfo {author}
  {\bibfnamefont {M.}~\bibnamefont {Tenje}},\ }\href@noop {} {\bibfield
  {journal} {\bibinfo  {journal} {Rep. Prog. Phys}\ }\textbf {\bibinfo {volume}
  {74}},\ \bibinfo {pages} {036101} (\bibinfo {year} {2011})}\BibitemShut
  {NoStop}%
\bibitem [{\citenamefont {Torres}\ \emph {et~al.}(2013)\citenamefont {Torres},
  \citenamefont {Meng}, \citenamefont {Ju}, \citenamefont {Zhao}, \citenamefont
  {Blair}, \citenamefont {Liu}, \citenamefont {Chao}, \citenamefont
  {Martyniuk}, \citenamefont {Roch-Jeune}, \citenamefont {Flaminio},\ and\
  \citenamefont {Michel}}]{Torres:2013ir}%
  \BibitemOpen
  \bibfield  {author} {\bibinfo {author} {\bibfnamefont {F.~A.}\ \bibnamefont
  {Torres}}, \bibinfo {author} {\bibfnamefont {P.}~\bibnamefont {Meng}},
  \bibinfo {author} {\bibfnamefont {L.}~\bibnamefont {Ju}}, \bibinfo {author}
  {\bibfnamefont {C.}~\bibnamefont {Zhao}}, \bibinfo {author} {\bibfnamefont
  {D.~G.}\ \bibnamefont {Blair}}, \bibinfo {author} {\bibfnamefont {K.~Y.}\
  \bibnamefont {Liu}}, \bibinfo {author} {\bibfnamefont {S.}~\bibnamefont
  {Chao}}, \bibinfo {author} {\bibfnamefont {M.}~\bibnamefont {Martyniuk}},
  \bibinfo {author} {\bibfnamefont {I.}~\bibnamefont {Roch-Jeune}}, \bibinfo
  {author} {\bibfnamefont {R.}~\bibnamefont {Flaminio}}, \ and\ \bibinfo
  {author} {\bibfnamefont {C.}~\bibnamefont {Michel}},\ }\href@noop {}
  {\bibfield  {journal} {\bibinfo  {journal} {Journal of Applied Physics}\
  }\textbf {\bibinfo {volume} {114}},\ \bibinfo {pages} {014506} (\bibinfo
  {year} {2013})}\BibitemShut {NoStop}%
\bibitem [{\citenamefont {Chang}\ \emph {et~al.}(2012)\citenamefont {Chang},
  \citenamefont {Banishev}, \citenamefont {Castillo-Garza}, \citenamefont
  {Klimchitskaya}, \citenamefont {Mostepanenko},\ and\ \citenamefont
  {Mohideen}}]{2012PhRvB..85p5443C}%
  \BibitemOpen
  \bibfield  {author} {\bibinfo {author} {\bibfnamefont {C.~C.}\ \bibnamefont
  {Chang}}, \bibinfo {author} {\bibfnamefont {A.~A.}\ \bibnamefont {Banishev}},
  \bibinfo {author} {\bibfnamefont {R.}~\bibnamefont {Castillo-Garza}},
  \bibinfo {author} {\bibfnamefont {G.~L.}\ \bibnamefont {Klimchitskaya}},
  \bibinfo {author} {\bibfnamefont {V.~M.}\ \bibnamefont {Mostepanenko}}, \
  and\ \bibinfo {author} {\bibfnamefont {U.}~\bibnamefont {Mohideen}},\
  }\href@noop {} {\bibfield  {journal} {\bibinfo  {journal} {Physical Review
  B}\ }\textbf {\bibinfo {volume} {85}},\ \bibinfo {pages} {165443} (\bibinfo
  {year} {2012})}\BibitemShut {NoStop}%
\bibitem [{\citenamefont {Safavi-Naeini}\ \emph {et~al.}(2013)\citenamefont
  {Safavi-Naeini}, \citenamefont {Chan}, \citenamefont {Hill}, \citenamefont
  {Gr{\"o}blacher}, \citenamefont {Miao}, \citenamefont {Chen}, \citenamefont
  {Aspelmeyer},\ and\ \citenamefont {Painter}}]{SafaviNaeini:2013cr}%
  \BibitemOpen
  \bibfield  {author} {\bibinfo {author} {\bibfnamefont {A.~H.}\ \bibnamefont
  {Safavi-Naeini}}, \bibinfo {author} {\bibfnamefont {J.}~\bibnamefont {Chan}},
  \bibinfo {author} {\bibfnamefont {J.~T.}\ \bibnamefont {Hill}}, \bibinfo
  {author} {\bibfnamefont {S.}~\bibnamefont {Gr{\"o}blacher}}, \bibinfo
  {author} {\bibfnamefont {H.}~\bibnamefont {Miao}}, \bibinfo {author}
  {\bibfnamefont {Y.}~\bibnamefont {Chen}}, \bibinfo {author} {\bibfnamefont
  {M.}~\bibnamefont {Aspelmeyer}}, \ and\ \bibinfo {author} {\bibfnamefont
  {O.}~\bibnamefont {Painter}},\ }\href@noop {} {\bibfield  {journal} {\bibinfo
   {journal} {New Journal of Physics}\ }\textbf {\bibinfo {volume} {15}},\
  \bibinfo {pages} {035007} (\bibinfo {year} {2013})}\BibitemShut {NoStop}%
\bibitem [{\citenamefont {Purdy}\ \emph {et~al.}(2013)\citenamefont {Purdy},
  \citenamefont {Yu}, \citenamefont {Peterson}, \citenamefont {Kampel},\ and\
  \citenamefont {Regal}}]{Purdy:2013wf}%
  \BibitemOpen
  \bibfield  {author} {\bibinfo {author} {\bibfnamefont {T.~P.}\ \bibnamefont
  {Purdy}}, \bibinfo {author} {\bibfnamefont {P.~L.}\ \bibnamefont {Yu}},
  \bibinfo {author} {\bibfnamefont {R.~W.}\ \bibnamefont {Peterson}}, \bibinfo
  {author} {\bibfnamefont {N.~S.}\ \bibnamefont {Kampel}}, \ and\ \bibinfo
  {author} {\bibfnamefont {C.~A.}\ \bibnamefont {Regal}},\ }\href@noop {}
  {\bibfield  {journal} {\bibinfo  {journal} {Physical Review X}\ }\textbf
  {\bibinfo {volume} {3}},\ \bibinfo {pages} {031012} (\bibinfo {year}
  {2013})}\BibitemShut {NoStop}%
\bibitem [{\citenamefont {Driggers}\ \emph {et~al.}(2012)\citenamefont
  {Driggers}, \citenamefont {Harms},\ and\ \citenamefont
  {Adhikari}}]{NN:subtract2012}%
  \BibitemOpen
  \bibfield  {author} {\bibinfo {author} {\bibfnamefont {J.~C.}\ \bibnamefont
  {Driggers}}, \bibinfo {author} {\bibfnamefont {J.}~\bibnamefont {Harms}}, \
  and\ \bibinfo {author} {\bibfnamefont {R.~X.}\ \bibnamefont {Adhikari}},\
  }\href {\doibase 10.1103/PhysRevD.86.102001} {\bibfield  {journal} {\bibinfo
  {journal} {Phys. Rev. D}\ }\textbf {\bibinfo {volume} {86}},\ \bibinfo
  {pages} {102001} (\bibinfo {year} {2012})}\BibitemShut {NoStop}%
\bibitem [{\citenamefont {{Creighton}}(2008)}]{Teviet:2008}%
  \BibitemOpen
  \bibfield  {author} {\bibinfo {author} {\bibfnamefont {T.}~\bibnamefont
  {{Creighton}}},\ }\href {\doibase 10.1088/0264-9381/25/12/125011} {\bibfield
  {journal} {\bibinfo  {journal} {Classical and Quantum Gravity}\ }\textbf
  {\bibinfo {volume} {25}},\ \bibinfo {eid} {125011} (\bibinfo {year}
  {2008})},\ \Eprint {http://arxiv.org/abs/gr-qc/0007050} {gr-qc/0007050}
  \BibitemShut {NoStop}%
\bibitem [{\citenamefont {Mazo}(1959)}]{Mazo195957}%
  \BibitemOpen
  \bibfield  {author} {\bibinfo {author} {\bibfnamefont {R.}~\bibnamefont
  {Mazo}},\ }\href {\doibase http://dx.doi.org/10.1016/S0031-8914(59)91235-2}
  {\bibfield  {journal} {\bibinfo  {journal} {Physica}\ }\textbf {\bibinfo
  {volume} {25}},\ \bibinfo {pages} {57 } (\bibinfo {year} {1959})}\BibitemShut
  {NoStop}%
\bibitem [{\citenamefont {Braginsky}\ \emph {et~al.}(1999)\citenamefont
  {Braginsky}, \citenamefont {Gorodetsky},\ and\ \citenamefont
  {Vyatchanin}}]{BGV1999}%
  \BibitemOpen
  \bibfield  {author} {\bibinfo {author} {\bibfnamefont {V.~B.}\ \bibnamefont
  {Braginsky}}, \bibinfo {author} {\bibfnamefont {M.~L.}\ \bibnamefont
  {Gorodetsky}}, \ and\ \bibinfo {author} {\bibfnamefont {S.~P.}\ \bibnamefont
  {Vyatchanin}},\ }\href@noop {} {\bibfield  {journal} {\bibinfo  {journal}
  {Phys.~Lett.~A}\ }\textbf {\bibinfo {volume} {264}},\ \bibinfo {pages} {1}
  (\bibinfo {year} {1999})}\BibitemShut {NoStop}%
\bibitem [{\citenamefont {Braginsky}\ \emph {et~al.}(2000)\citenamefont
  {Braginsky}, \citenamefont {Gorodetsky},\ and\ \citenamefont
  {Vyatchanin}}]{BGV2000}%
  \BibitemOpen
  \bibfield  {author} {\bibinfo {author} {\bibfnamefont {V.~B.}\ \bibnamefont
  {Braginsky}}, \bibinfo {author} {\bibfnamefont {M.~L.}\ \bibnamefont
  {Gorodetsky}}, \ and\ \bibinfo {author} {\bibfnamefont {S.~P.}\ \bibnamefont
  {Vyatchanin}},\ }\href@noop {} {\bibfield  {journal} {\bibinfo  {journal}
  {Phys.~Lett.~A}\ }\textbf {\bibinfo {volume} {271}},\ \bibinfo {pages} {303}
  (\bibinfo {year} {2000})}\BibitemShut {NoStop}%
\bibitem [{\citenamefont {Levin}(2008)}]{Levin20081941}%
  \BibitemOpen
  \bibfield  {author} {\bibinfo {author} {\bibfnamefont {Y.}~\bibnamefont
  {Levin}},\ }\href {\doibase http://dx.doi.org/10.1016/j.physleta.2007.11.007}
  {\bibfield  {journal} {\bibinfo  {journal} {Physics Letters A}\ }\textbf
  {\bibinfo {volume} {372}},\ \bibinfo {pages} {1941 } (\bibinfo {year}
  {2008})}\BibitemShut {NoStop}%
\bibitem [{\citenamefont {{Clerk}}\ \emph {et~al.}(2010)\citenamefont
  {{Clerk}}, \citenamefont {{Devoret}}, \citenamefont {{Girvin}}, \citenamefont
  {{Marquardt}},\ and\ \citenamefont {{Schoelkopf}}}]{Clerk:RMP2010}%
  \BibitemOpen
  \bibfield  {author} {\bibinfo {author} {\bibfnamefont {A.~A.}\ \bibnamefont
  {{Clerk}}}, \bibinfo {author} {\bibfnamefont {M.~H.}\ \bibnamefont
  {{Devoret}}}, \bibinfo {author} {\bibfnamefont {S.~M.}\ \bibnamefont
  {{Girvin}}}, \bibinfo {author} {\bibfnamefont {F.}~\bibnamefont
  {{Marquardt}}}, \ and\ \bibinfo {author} {\bibfnamefont {R.~J.}\ \bibnamefont
  {{Schoelkopf}}},\ }\href {\doibase 10.1103/RevModPhys.82.1155} {\bibfield
  {journal} {\bibinfo  {journal} {Reviews of Modern Physics}\ }\textbf
  {\bibinfo {volume} {82}},\ \bibinfo {pages} {1155} (\bibinfo {year}
  {2010})},\ \Eprint {http://arxiv.org/abs/0810.4729} {arXiv:0810.4729
  [cond-mat.mes-hall]} \BibitemShut {NoStop}%
\bibitem [{\citenamefont {Chen}(2013{\natexlab{b}})}]{Yanbei:MQMreview}%
  \BibitemOpen
  \bibfield  {author} {\bibinfo {author} {\bibfnamefont {Y.}~\bibnamefont
  {Chen}},\ }\href@noop {} {\bibfield  {journal} {\bibinfo  {journal} {Journal
  of Physics B: Atomic, Molecular and Optical Physics}\ }\textbf {\bibinfo
  {volume} {46}},\ \bibinfo {pages} {104001} (\bibinfo {year}
  {2013}{\natexlab{b}})}\BibitemShut {NoStop}%
\bibitem [{\citenamefont {Morrison}(1998)}]{morrison1998grounding}%
  \BibitemOpen
  \bibfield  {author} {\bibinfo {author} {\bibfnamefont {R.}~\bibnamefont
  {Morrison}},\ }\href {http://books.google.com/books?id=\_TFTAAAAMAAJ} {\emph
  {\bibinfo {title} {Grounding and shielding techniques}}},\ Wiley-Interscience
  publication\ (\bibinfo  {publisher} {Wiley},\ \bibinfo {year}
  {1998})\BibitemShut {NoStop}%
\end{thebibliography}%

\end{document}